\documentclass[fleqn,usenatbib]{mnras}

\usepackage{newtxtext,newtxmath}

\usepackage[T1]{fontenc}
\usepackage{ae,aecompl}

\usepackage{graphicx}	
\usepackage{amsmath}	
\usepackage{amssymb}	

\newcommand{\sparcfire}{\textsc{SpArc\-FiRe}}
\newcommand{\spotter}{\textsc{Spiral\-Spotter}}
\usepackage[usenames,dvipsnames]{xcolor}

\title[Galaxy Zoo and \sparcfire{}]{Galaxy Zoo and \sparcfire{}: Constraints on spiral arm formation mechanisms from spiral arm number and pitch angles}

\author[Hart et al.]{Ross E. Hart,$^{1}$\thanks{E-mail: ross.hart@nottingham.ac.uk}
Steven P. Bamford,$^{1}$ Wayne B. Hayes,$^{2}$ Carolin N. Cardamone,$^{3}$ \newauthor William C. Keel,$^{4}$ Sandor J. Kruk,$^{5}$ Chris J. Lintott,$^{5}$ Karen L. Masters,$^{6}$ \newauthor Brooke D. Simmons,$^{7}$ Rebecca J. Smethurst$^{1}$  \\
$^{1}$School of Physics \& Astronomy, The University of Nottingham, University Park, Nottingham NG7 2RD, UK\\
$^{2}$University of California, Irvine, CA 92697-3435, USA\\
$^{3}$Math and Science Department, Wheelock College, 200 The Riverway, Boston, MA 02215, USA\\
$^{4}$Department of Physics and Astronomy, University of Alabama, Box 870324, Tuscaloosa, AL 35487, USA\\
$^{5}$Oxford Astrophysics, Denys Wilkinson Building, Keble Road, Oxford OX1 3RH, UK\\
$^{6}$Institute for Cosmology and Gravitation, University of Portsmouth, Dennis Sciama Building, Portsmouth PO1 3FX, UK\\
$^{7}$Center for Astrophysics and Space Sciences (CASS), Department of Physics, University of California, San Diego, CA 92093, USA\\
}

\date{Accepted XXX. Received YYY; in original form ZZZ}

\pubyear{2017}

\begin{document}
\label{firstpage}
\pagerange{\pageref{firstpage}--\pageref{lastpage}}
\maketitle

\begin{abstract}
In this paper we study the morphological properties of spiral galaxies, including measurements of spiral arm number and pitch angle. Using Galaxy Zoo 2, a stellar mass-complete sample of 6,222 SDSS spiral galaxies is selected. We use the machine vision algorithm \sparcfire{} to identify spiral arm features and measure their associated geometries. A support vector machine classifier is employed to identify reliable spiral features, with which we are able to estimate pitch angles for half of our sample.  We use these machine measurements to calibrate visual estimates of arm tightness, and hence estimate pitch angles for our entire sample. The properties of spiral arms are compared with respect to various galaxy properties. The star formation properties of galaxies vary significantly with arm number, but not pitch angle. We find that galaxies hosting strong bars have spiral arms substantially (4-6\textdegree{}) looser than unbarred galaxies.  Accounting for this, spiral arms associated with many-arm structures are looser (by 2\textdegree{}) than those in two-arm galaxies.  In contrast to this average trend, galaxies with greater bulge-to-total stellar mass ratios display both fewer and looser spiral arms. This effect is primarily driven by the galaxy disc, such that galaxies with more massive discs contain more spiral arms with tighter pitch angles. This implies that galaxy central mass concentration is not the dominant cause of pitch angle and arm number variations between galaxies, which in turn suggests that not all spiral arms are governed by classical density waves or modal theories.
\end{abstract}

\begin{keywords}
galaxies: general -- galaxies: spiral -- galaxies: structure -- methods: data analysis
\end{keywords}



\section{Introduction}
\label{sec:introduction}

Spiral arms are morphological features observed in the majority of low-redshift galaxies. Although low-mass galaxies tend to have irregular structures \citep{Binggeli_85}, spiral arms are common in massive galaxies with extended discs \citet{Nair_10,Lintott_11,Willett_13}. They are sites of enhanced gas density \citep{Grabelsky_87,Engargiola_03,Sanchez_Menguiano_17}, dust \citep{Sandage_61,Holwerda_05} and star formation \citep{Calzetti_05,Grosbol_12} that form beautiful, sweeping patterns in galaxy discs. Describing a galaxy's morphology as `spiral' is an imprecise description, as it encompasses a wide range of types of spiral structure. Instead, spiral structure can be further described as one of three types -- grand design, many-arm or flocculent \citet{Elmegreen_82}. Grand design spirals have two distinct spiral arms, and many-arm patterns have multiple global spiral arms extending to the edges of galaxy discs. Flocculent spirals also have multiple arms, but the arms are much weaker and less distinct. These structures have a variety of formation mechanisms, but evaluating their significance requires a more quantitative approach. Two useful geometric parameters that we can derive for spirals are the number of spiral arms and the pitch angle. The former is the number of clear, well-defined spiral features, while the latter is a measure of how tightly-wrapped the arms are.

The form of the equation describing the path of a spiral arm offers information on the underlying processes that are responsible for the spiral arm. Logarithmic spiral arms are usually indicative of density waves, whereas material arms can usually be described by a hyperbolic function. Such a function is directly proportional to the galaxy rotation velocity, as material arms rotate rigidly with the galaxy disc \citep{Kennicutt_81}. It has been demonstrated that most spiral arms can be well described by log spiral arcs \citep{Rots_74,Rots_75,Boeshaar_77,Kennicutt_82,Davis_14}, which in turn suggests that spiral arms are density enhancements due to the presence of density waves or other similar mechanisms, and are not material in nature \citep{Seigar_98a,Seigar_98b,Donghia_13}.

The geometry of spiral patterns is well-studied in local galaxies, in terms of both spiral arm number and pitch angle. Although early studies of spirals were restricted to small samples of a few hundred galaxies (e.g. \citealt{Elmegreen_82,Elmegreen_82b,Elmegreen_89,Ann_13}), Galaxy Zoo classifications have recently allowed for the study of statistically complete samples of spirals an order of magnitude larger \citep{Hart_16,Hart_17}. Two-armed, grand design structures are more prevalent in high density environments \citep{Elmegreen_82b,Hart_16}, more likely to occur in the presence of a strong bar \citep{Elmegreen_82,Elmegreen_89}, and have more dust obscured star formation \citep{Hart_17}, irrespective of galaxy stellar mass. These are significant clues that the processes responsible for observed grand design spiral arms differ from those that lead to many-armed, flocculent patterns, and that bars and local environment may play a role in triggering a two-arm pattern in galaxies \citep{Elmegreen_83,Dobbs_10,Semczuk_17}.

Despite the advances that Galaxy Zoo has made in terms of the study of galaxy morphology in large samples of low-redshift galaxies, measures of spiral arm pitch angles for such large samples remain elusive. Instead, much smaller samples have been used, with typically $\lesssim$100 galaxies (e.g. \citealt{Seigar_05,Seigar_06,Martinez_Garcia_12,Savchenko_13}). These studies of local galaxies reveal interesting trends relating spiral arm geometry with fundamental galaxy properties. It has been established that spiral arm pitch angle is strongly correlated with the rotation properties of galaxies: galaxies with higher rotation velocities have more loosely wound spiral arms \citep{Kennicutt_81}, and pitch angles are even more closely correlated to galaxy rotation curves \citep{Seigar_05,Seigar_06}. These results imply that the underlying mass distribution of galaxies directly affects the shapes of spiral arms \citep{Seigar_08,Seigar_14,Berrier_13}, explaining why galaxies with more tightly wound arms are often associated with greater central mass concentrations \citep{Hubble_26}. Although this link has been clearly established, it has only been observed in small samples of nearby grand design spirals. Galaxies which display different types of spiral structure could have a different explanation. Simulations show that many-armed structures should have higher pitch angles \citep{Donghia_13,Grand_13} and there is evidence that weaker, multi-arm spiral patterns are more open \citep{Puerari_92}. In these simulated galaxies, spiral arms wind up, becoming tighter over time, meaning that pitch angle may also indicate the age of the arm feature \citep{Perez_Villegas_12,Grand_13}. Strong bars can also influence the pitch angles of spiral galaxies \citep{Athanassoula_09a,Martinez_Garcia_12,Baba_15}. If a bar is strong enough and extends beyond the galaxy corotation radius, the nature of spiral arms could change completely, from density waves to material arms amplified at the end of the bar \citep{Roca_Fabrega_13}. Another factor to consider is galaxy-galaxy interactions -- such interactions can morphologically disturb galaxies (e.g. \citealt{Ellison_10,Kaviraj_14,Patton_16}), leading to more open arms in galaxy-galaxy separations $\lesssim100$kpc \citep{Casteels_13}. Testing the effects that the aforementioned processes have on the structure of spiral arms requires statistically complete samples of spiral galaxies with measured arm pitch angles.

Automated methods offer an interesting prospect for measuring pitch angles in large galaxy samples. Although they still cannot measure overall morphological parameters to the same level as human inspection, they do give an opportunity to study spiral arm geometries in more detail (e.g. \citealt{Considere_88,Puerari_92,Saraiva_94,Rix_95,Davis_12}). In this paper, we study the geometry of spiral arms in a large sample of spiral galaxies combining visual classification statistics from Galaxy Zoo with an automated method of detecting spiral arms called \sparcfire{} (see \citealt{Davis_14} for a full description).

This paper is organised as follows. In Section 2, the sample selection and galaxy data are described. This includes a description of how spiral arm pitch angles are derived from \sparcfire{}. In Section 3, spiral arm pitch angles are studied as a function of other galaxy properties, namely spiral arm number, bar strength, central mass concentration, and star formation rate (SFR). The results and their implications with respect to relevant theoretical and observational literature are discussed in Section 5. Our conclusions are summarised in Section 6.

This paper assumes a flat cosmology with $\Omega_\mathrm{m}=0.3$ and $H_0=70 \mathrm{km s^{-1} \, Mpc^{-1}}$.

\section{Data}
\label{sec:data}

\subsection{Galaxy properties}
\label{sec:galaxy_properties_and_sample_selection}

All visual galaxy morphological information is obtained from the public data release of Galaxy Zoo 2 (GZ2; \citealt{Willett_13}). The questions that are considered concern whether spiral arms are present, how  tightly wound the spiral arms are and how many spiral arms there are in a galaxy. As these are `third branch' questions in GZ2 (see \citealt{Willett_13} for more details about the GZ2 question tree) with multiple answers, we use the debiased statistics from \citet{Hart_16}\footnote{GZ2 morphological measurements are available from data.galaxyzoo.org}, which are consistent classifications free of redshift-dependent bias caused by image degradation at higher redshift. These are biases in the galaxy population  Supplementary morphological data is included from Galaxy Zoo 1 (GZ1; \citealt{Lintott_11}). The galaxies classified in GZ2 were taken from the SDSS main galaxy sample, which is an $r$-band selected sample of galaxies in the legacy imaging area targeted for spectroscopic follow-up \citep{Strauss_02}. The \citet{Hart_16} sample contains all well-resolved galaxies in SDSS DR7 \citep{Abazijian_09} to a limiting magnitude of $m_r \leq 17.0$. In this paper, we consider galaxies classified in the normal-depth (single-epoch) DR7 imaging with spectroscopic redshifts. Spectroscopic redshifts are required for galaxies to have morphological data corrected for redshift-dependent classification bias (see \citealt{Bamford_09} and \citealt{Hart_16}) and accurate measurements of rest frame photometry. Rest frame optical photometry and redshifts are obtained from the SDSS DR7 catalogue (see \citealt{Bamford_09} for a detailed description). Galaxy stellar masses are obtained from the \citet{Mendel_14} SED fits. As we expect most visually classified spirals to be two-component bulge-disc systems, we use the bulge+disc masses from \citet{Mendel_14}. We note that these masses are largely consistent with those used in \citet{Hart_17}, which were obtained from the SDSS+WISE catalogue of \citet{Chang_15}, with a small offset of $-0.07$ dex and scatter of $0.10$ dex. 

Galaxy UV fluxes are from the GALEX GR6 catalogue \citep{Martin_05}, which are included in the NASA Sloan Atlas \citep{Blanton_11}. Mid-IR fluxes are from the AllWISE catalogue \citep{Wright_10}, and obtained from the reduced photometry of \citet{Chang_15}. Details of the matching procedure are included in \citet{Hart_17}. Star formation rates (SFRs) are calibrated using the standard conversions of \citet{Buat_08,Buat_11} for UV fluxes and \citet{Jarrett_13} for mid-IR fluxes. The conversion factors are detailed in \citet{Hart_17}.

\subsection{Sample selection}
\label{sec:sample_selection}

\subsubsection{Spiral galaxy selection}
\label{sec:spiral_selection}

In Galaxy Zoo 2, each response to a question is assigned a value of $p$, which takes the value $0 \leq p \leq 1$ (e.g. if half of respondents said a feature was present, this would mean $p=0.5$). Spiral galaxies are selected using the same criteria of \citet{Hart_17}, by selecting galaxies with $p_\mathrm{features} \cdot p_\mathrm{not \, edge \, on} \cdot p_\mathrm{spiral} \geq 0.5$. We also limit our sample to galaxies with redshift $0.02 < z \leq 0.055$ to select the most reliably identified spiral galaxies, and to ensure the greatest reliability in the \citet{Mendel_14} bulge-disc mass measurements. We ensure all of our galaxies are relatively face-on by using a cut of $(b/a)_g>0.4$, where $a$ and $b$ are the isophotal semi-major and minor axes in the $g$-band. This corresponds to an inclination $i<70$\textdegree{} for disc thickness $q=0.22$, e.g. \citealt{Unterborn_08}. Galaxy Zoo statistics can reliably identify bars \citep{Masters_11} and spiral arms \citep{Hart_17} for galaxies more face-on than this threshold.

\subsubsection{Spiral arm number}
\label{sec:arm_number}

The spiral arm number in a galaxy can be identified in two ways. The first is an absolute quantity, $m$, as used in \citet{Hart_16}. It is defined as the response to the spiral arm number question in GZ2 which has the highest vote fraction. It can take five discrete values: 1, 2, 3, 4 or 5+ spiral arms. It is sometimes desirable to define a more continuous statistic for measuring arm number, which does not have discrete values. We therefore define $m_\mathrm{avg}$, which is the average of all of the arm number responses, given by
\begin{equation}
    \label{eq:m_wtd_avg}
    m_\mathrm{avg} = \sum_{m=1}^{5} m p_\mathrm{m},
\end{equation}
where $m$ is the value assigned to each response in turn (1, 2, 3, 4 or 5) and $p_m$ is the fraction of votes for that response. The statistic can take any value in the range 1-5, where $m_\mathrm{avg}=1$ means all volunteers said a galaxy had one spiral arm, and $m_\mathrm{avg}=5$ means all classifiers said a galaxy had 5+ spiral arms. 

\subsubsection{The presence of bars}
\label{sec:bar_presence}

The presence of bars in our galaxies is measured using the response to the `is there a bar?' question in GZ2. A continuous statistic for this purpose is the quantity $p_\mathrm{bar}$, defined as the fraction of responses that said a bar was present in a galaxy. The bar question has been shown to be an effective method for measuring not only the presence of bars, but also the strength of bars in galaxies \citep{Skibba_12}. In some of our analysis, we would like to compare galaxy properties irrespective of the presence of bars. We use cuts of $p_\mathrm{bar}>$0.5 to define subsamples of strongly barred galaxies, $0.2 < p_\mathrm{bar} \leq 0.5$  to define weakly barred galaxies and $p_\mathrm{bar} \leq 0.2$ to define unbarred galaxies \citep{Masters_11,Skibba_12} in our spiral sample. The numbers of galaxies in subsamples made using these cuts are detailed in Table~\ref{table:galaxy_parameters}.

\subsubsection{Stellar mass completeness}
\label{sec:stellar_mass_completeness}

\begin{figure}
    \includegraphics[width=0.45\textwidth]{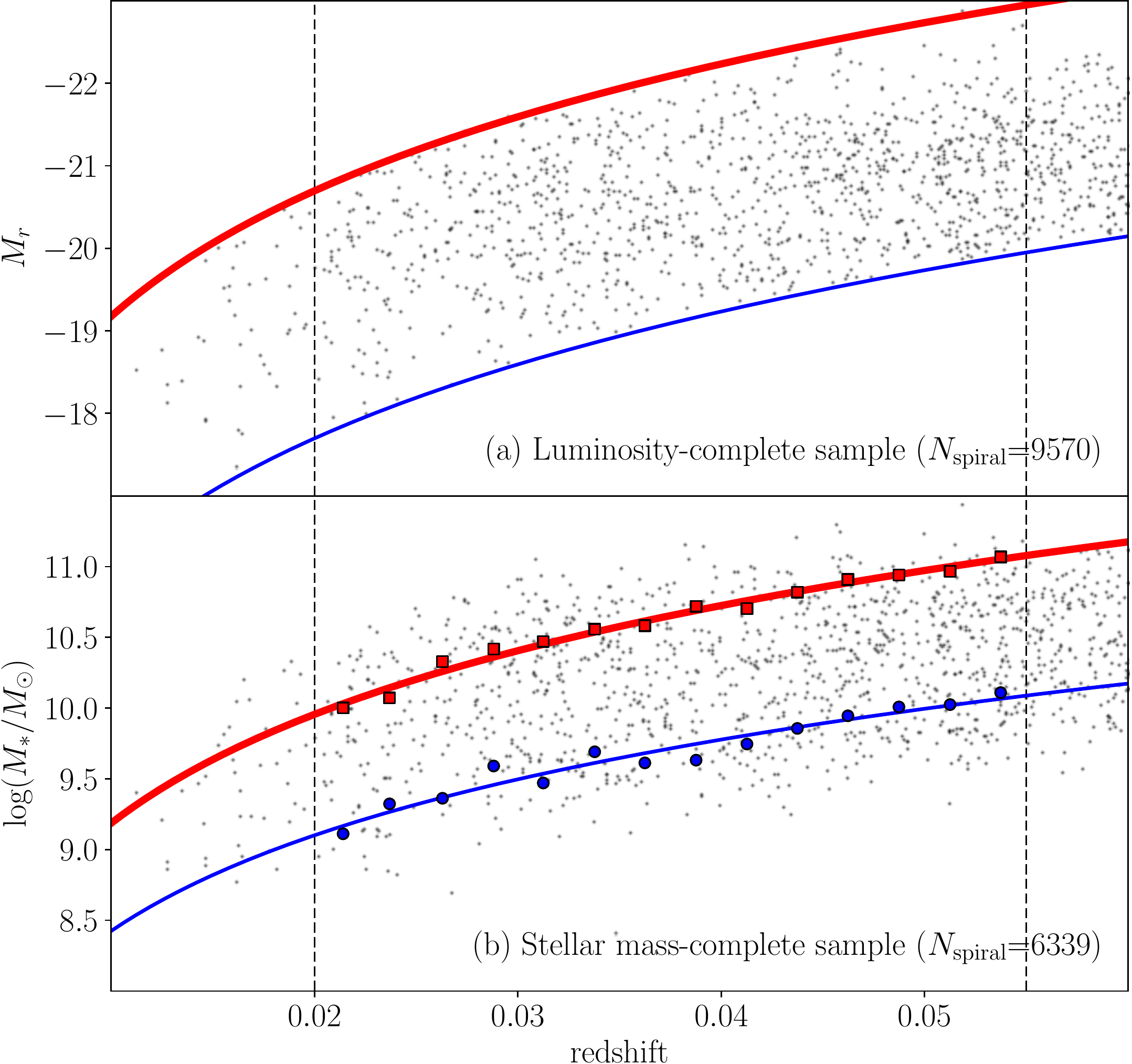}
    \caption{(a) Galaxy redshift vs. absolute $r$-band magnitude. The black points show individual galaxies. The thinner blue line indicates the faint magnitude limit ($m_r$=17), and the thicker red line shows the bright magnitude limit ($m_r=14$). (b) Galaxy redshift vs. total galaxy stellar mass \citep{Mendel_14}. The black points show individual galaxies. The red squares show the upper mass limits in bins of $\Delta z$=0.0025, and the blue circles show the lower mass limits. The vertical dashed black lines show the redshift limits of $0.02 < z \leq 0.055$. The best fit lines to the red squares is shown by the thicker red line, and the best fit to the blue circles is shown by the thinner blue line. In both panels, a subset of 4,000 galaxies are shown for clarity.}
    \label{fig:sample_limiting}
\end{figure}

In order to study galaxy properties in a representative manner, we wish to define a sample of spiral galaxies complete in stellar mass. The \citet{Mendel_14} catalogue contains galaxies with $14 < m_r \leq 17.77$. In GZ2, a faint magnitude limit of $m_r \leq 17$ is also applied. All galaxies with $0.02 < z \leq 0.055$ with magnitudes $14 < m_r \leq 17$ are included in the \textit{flux-complete sample}. The thin blue line in Fig.~\ref{fig:sample_limiting}a shows the faint magnitude limit as a function of redshift, and the thicker red line shows the bright end limit imposed in \citet{Mendel_14}. In total, there are 12042 spiral galaxies with $0.02 < z \leq 0.055$ in the \textit{flux-complete sample}. 

In order to define a sample complete in stellar mass, the stellar mass completeness limits are computed with redshift. We follow the method of \citet[][later used in a low-redshift SDSS sample in \citealt{Weigel_16}]{Pozzetti_10} to define the stellar mass completeness and the steps are outlined below. Spiral galaxies were binned by redshift in bins of $\Delta z$=0.0025. Each galaxy in a bin was then assigned a limiting mass, $M_\mathrm{*,lim}$, defined as the mass the galaxy would have if its luminosity was that of the faint luminosity-limit of the survey at the galaxy's redshift and it had the same mass-to-light ratio, $M_*/L_r$. We then selected the faintest 20\% of galaxies in the bin. The stellar mass completeness limit, $M_\mathrm{*,lower}$, was measured as the mass below which lay 95\% of the $M_\mathrm{*,lim}$ values of this faintest 20\% subsample. This was computed for each bin in turn, and they are shown by the blue circles in Fig.~\ref{fig:sample_limiting}b. As our sample also includes a bright magnitude limit, we compute the upper mass limit, $M_\mathrm{*,upper}$, by calculating the maximum mass galaxies could have, $M_\mathrm{*,lim \, max}$, and measuring the mass above which 95 per cent of the limiting masses of the 20 percent brightest galaxies in each bin lay. These are shown by the red squares in Fig.~\ref{fig:sample_limiting}b. The limiting masses with redshift were then measured by fitting a log curve to the upper and lower mass limits, and they take the form \begin{equation}
\label{eq:mass_lower}
\log(M_\mathrm{*,lower}/M_\mathrm{\odot}) = 2.07 (\pm 0.15) \log(z) + 12.64 (\pm 0.21) ,
\end{equation} and \begin{equation}
\label{eq:mass_upper}
\log(M_\mathrm{*,upper}/M_\mathrm{\odot})  = 2.45 (\pm 0.08) \log(z) + 14.05 (\pm 0.12).
\end{equation} The $\pm$ values indicate the error in each fitted parameter, obtained from the covariance matrix. These lower and upper mass limits are shown by the thin blue line and thicker red line in Fig.~\ref{fig:sample_limiting}b. In total there are 6339 galaxies in the \textit{stellar mass-complete sample} between these limits.

In order to sample fairly for all stellar masses, a volume correction is applied. This means the \textit{stellar mass-complete sample} can mimic a \textit{stellar mass-limited sample}. For each galaxy, the maximum volume is calculated using the upper and lower redshift bounds where a galaxy with its stellar mass could fall within the stellar-mass completeness limits defined above. Each galaxy is then assigned a $1/V_\mathrm{max}$ weighting. We remove any galaxies that lie in a very small volume, and thus having large $1/V_\mathrm{max}$ corrections, by only selecting galaxies in $9.45 < \log(M_*/M_\odot) \leq 11.05$ (corresponding to $1/V_\mathrm{max} \leq 10$). In total, 117 (1.8 \%) of the galaxies were removed for this reason, leaving  a final \textit{stellar mass-complete sample} of 6222 spiral galaxies. These samples are further subdivided into spiral galaxies with different arm numbers and bar probabilities. The number of spiral galaxies in each subsample of the \textit{stellar mass-complete sample} is given in $N_\mathrm{gal}$ of Table~\ref{table:galaxy_parameters}, and the median, 16th and 84th percentile stellar masses is given in the $\log(M_*/M_\mathrm\odot)$ column of Table~\ref{table:galaxy_parameters}.

\begin{table*}
\caption{Galaxy sample parameters for all of our samples of galaxies. For the $N$ columns, the first number indicates the total number of galaxies, and the bracketed number indicates the fraction of those galaxies with at least one good spiral arm in \sparcfire{} (see Sec.~\ref{sec:identifying_arms}). In the stellar mass columns, the first number indicates the median stellar mass, and the bracketed values are the 16th and 84th percentiles. The * next to the $m=4$ (all) sample indicates that it is the sample that was used as the reference sample for matching in stellar mass.}
\begin{tabular}{ccccc}

\hline
 subsample                                & $N$ (all)    & $\log[M_*/M_\odot]$ (all)   & $N$ ($M_*$-matched)   & $\log[M_*/M_\odot]$ ($M_*$-matched)   \\
\hline
 spiral (all)                             & 6222 (48.7\%) & 10.27 (9.89, 10.64)   & 4908 (48.2\%) & 10.26 (9.92, 10.58)   \\
 $m=1$ (all)                              & 243 (25.9\%)  & 10.18 (9.78, 10.57)   & 151 (20.5\%)  & 10.25 (9.92, 10.56)   \\
 $m=2$ (all)                              & 4014 (46.6\%) & 10.26 (9.88, 10.63)   & 3208 (45.6\%) & 10.26 (9.93, 10.58)   \\
 $m=3$ (all)                              & 1108 (56.9\%) & 10.32 (9.93, 10.66)   & 876 (57.5\%)  & 10.25 (9.91, 10.56)   \\
 $m=4$ (all)                              & 405 (54.6\%)  & 10.31 (9.93, 10.67)   & 337 (55.5\%)  & 10.26 (9.92, 10.56)   \\
 $m=5+$ (all)                             & 452 (53.8\%)  & 10.27 (9.88, 10.62)   & 336 (53.3\%)  & 10.24 (9.93, 10.57)   \\
 
\hline
 
 spiral ($p_\mathrm{bar} \leq 0.2$)       & 2237 (48.4\%) & 10.23 (9.89, 10.56)   & 1920 (47.6\%) & 10.26 (9.92, 10.57)   \\
 $m=1$ ($p_\mathrm{bar} \leq 0.2$)        & 135 (20.7\%)  & 10.18 (9.76, 10.54)   & 93 (16.1\%)   & 10.3 (9.98, 10.58)    \\
 $m=2$ ($p_\mathrm{bar} \leq 0.2$)        & 1034 (46.9\%) & 10.19 (9.87, 10.53)   & 879 (44.7\%)  & 10.25 (9.91, 10.57)   \\
 $m=3$ ($p_\mathrm{bar} \leq 0.2$)        & 570 (53.5\%)  & 10.27 (9.93, 10.58)   & 514 (54.1\%)  & 10.26 (9.93, 10.57)   \\
 $m=4$ ($p_\mathrm{bar} \leq 0.2$)*       & 221 (52.0\%)  & 10.26 (9.93, 10.59)   & 221 (52.0\%)  & 10.26 (9.93, 10.59)   \\
 $m=5+$ ($p_\mathrm{bar} \leq 0.2$)       & 277 (53.8\%)  & 10.25 (9.88, 10.58)   & 213 (53.1\%)  & 10.27 (9.93, 10.58)   \\
 
\hline 
 
 spiral ($0.2 < p_\mathrm{bar} \leq 0.5$) & 1858 (50.4\%) & 10.25 (9.89, 10.61)   & 1554 (50.3\%) & 10.25 (9.93, 10.56)   \\
 $m=1$ ($0.2 < p_\mathrm{bar} \leq 0.5$)  & 79 (27.8\%)   & 10.24 (9.81, 10.6)    & 43 (27.9\%)   & 10.24 (9.87, 10.5)    \\
 $m=2$ ($0.2 < p_\mathrm{bar} \leq 0.5$)  & 1226 (48.2\%) & 10.23 (9.89, 10.57)   & 1081 (47.5\%) & 10.26 (9.94, 10.57)   \\
 $m=3$ ($0.2 < p_\mathrm{bar} \leq 0.5$)  & 330 (60.0\%)  & 10.34 (9.94, 10.65)   & 256 (60.2\%)  & 10.27 (9.89, 10.56)   \\
 $m=4$ ($0.2 < p_\mathrm{bar} \leq 0.5$)  & 115 (59.1\%)  & 10.32 (9.9, 10.66)    & 88 (61.4\%)   & 10.26 (9.91, 10.51)   \\
 $m=5+$ ($0.2 < p_\mathrm{bar} \leq 0.5$) & 108 (53.7\%)  & 10.27 (9.91, 10.61)   & 86 (54.7\%)   & 10.21 (9.92, 10.52)   \\
 
\hline

 spiral ($p_\mathrm{bar} > 0.5$)          & 2127 (47.4\%) & 10.35 (9.89, 10.72)   & 1434 (46.7\%) & 10.26 (9.92, 10.6)    \\
 $m=1$ ($p_\mathrm{bar} > 0.5$)           & 29 (44.8\%)   & 10.17 (9.83, 10.82)   & 15 (26.7\%)   & 10.08 (9.89, 10.43)   \\
 $m=2$ ($p_\mathrm{bar} > 0.5$)           & 1754 (45.3\%) & 10.33 (9.88, 10.69)   & 1248 (44.6\%) & 10.27 (9.93, 10.6)    \\
 $m=3$ ($p_\mathrm{bar} > 0.5$)           & 208 (61.1\%)  & 10.51 (9.99, 10.8)    & 106 (67.9\%)  & 10.2 (9.9, 10.54)     \\
 $m=4$ ($p_\mathrm{bar} > 0.5$)           & 69 (55.1\%)   & 10.6 (9.96, 10.81)    & 28 (64.3\%)   & 10.18 (9.87, 10.51)   \\
 $m=5+$ ($p_\mathrm{bar} > 0.5$)          & 67 (53.7\%)   & 10.38 (9.85, 10.75)   & 37 (51.4\%)   & 10.24 (9.92, 10.66)   \\
\hline
\end{tabular}

\label{table:galaxy_parameters}

\end{table*}

\subsubsection{Matching in stellar mass}
\label{sec:mass_matching}

In Table~\ref{table:galaxy_parameters}, we see a small residual dependence of stellar mass on galaxy morphology, with galaxies with more spiral arms and stronger bars having greater stellar masses. In this paper, we wish to study properties of galaxies with respect to galaxy morphology only, with none of our results dependent on stellar mass. For this reason, we choose to match all of our subsamples in stellar mass to ensure there are no residual dependencies driving our results in later sections. The four-arm unbarred subsample is selected as the sample to match to (denoted by a * in Table~\ref{table:galaxy_parameters}) as it is the one with the fewest galaxies. We do note that there are actually fewer one-arm spirals, but these are a special case of galaxy, usually associated with mergers \citep{Casteels_13}, so contribute little to our analysis later in the paper.

To mass-match, we use a method that we call KDE-matching.\footnote{The source code for this method is publicly available at http://doi.org/10.5281/zenodo.815850} This process matches two distributions by a given statistic, which we call the \textit{reference sample} and the \textit{match sample}. In our example, the \textit{reference sample} is the $m=4$ subsample, and the \textit{match sample} is each of the other subsamples in turn. We convolve both samples with a Gaussian kernel with bandwidth optimised via five-fold cross-validation, resulting in a smoothed kernel density estimate (KDE) for both. The \textit{match} KDE is then divided by the \textit{reference} KDE, and each galaxy in the \textit{match sample} now has an associated probability. These probabilities are normalised so that the 95th percentile of all of the \textit{match} probabilities equals one. All galaxies in the top 5th percentile are set to probability $p=1$ and all other galaxies have probabilities $0<p<1$. Galaxies are then sampled from the \textit{match sample}, with probability, $p$ of being included in the final \textit{matched sample}. This process was used for all of the samples, and galaxies in these matched samples are in future referred to as \textit{mass-matched samples} or \textit{mass-matched subsamples}. The number of galaxies, and their associated stellar masses, for each \textit{mass-matched subsample} are given in the columns $N_\mathrm{gal}$ ($M_*$-matched) and $\log[M_*/M_\odot]$ ($M_*$-matched) of Table~\ref{table:galaxy_parameters}.

\subsection{Identifying spiral arms with \sparcfire{}}
\label{sec:identifying_arms}

Spiral arcs for all of our galaxies are measured using the automated method from \sparcfire{}\footnote{http://sparcfire.ics.uci.edu/} \citep{Davis_14}. Given an input image, \sparcfire{} identifies and fits logarithmic spiral arc structures. We apply the \sparcfire{} algorithm to the SDSS $r$-band images of our \textit{stellar mass-limited sample} of spiral galaxies. The method identifies several spiral arcs for each galaxy and only some of these correspond to true spiral arms. To correctly identify real spiral arms, \citet{Davis_14} compared their spiral arc statistics to those obtained from GZ2 and suggested selecting only arcs longer than 75 pixels. 

\subsubsection{\spotter{}}
\label{sec:spotter}

\begin{figure}
    \includegraphics[width=0.45\textwidth]{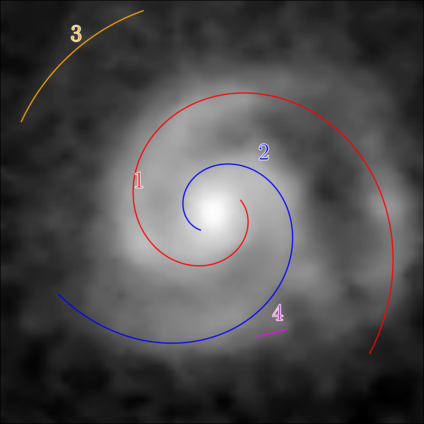}
    \caption{An example of a galaxy image presented to volunteers in \spotter{}. The greyscale image is the SDSS $r$-band image of the galaxy deprojected to face-on in \sparcfire{} (see \citealt{Davis_14} for details of this process). The coloured curves indicate where \sparcfire{} identified arcs in the image, each of which are assigned a number. Each arc was assigned a unique colour and number for volunteer classification.}
    \label{fig:spotter_subject}
\end{figure}

\begin{table}
\caption{Summary of the statistics identified by \spotter{}. People were asked whether arcs were good, poor, weak, extensions, junk or missing. The total number of classifications for each of these classes are shown in the second column. The third column shows the number of spiral arcs classed as one of these categories, using the category which had the greatest fraction of the votes.}

\begin{center}
\begin{tabular}{ccc}
\hline
 class       & $N_\mathrm{clicks}$   & $N_\mathrm{arcs}$   \\
\hline
 good      & 1088 (16.4\%)        & 244 (15.1\%)       \\
 poor      & 298 (4.5\%)          & 17 (1.1\%)         \\
 weak      & 713 (10.8\%)         & 85 (5.3\%)         \\
 extension & 669 (10.1\%)         & 104 (6.4\%)        \\
 junk      & 1175 (17.8\%)        & 190 (11.8\%)       \\
 missing   & 2673 (40.4\%)        & 678 (41.9\%)       \\
\hline
\end{tabular}
\end{center}
\label{table:spotter}
\end{table}

In this paper, we explore a more robust method of identifying arcs that correspond to real spiral arms, rather than simply relying on a single length cut. We aim to use a number of parameters to identify whether a given \sparcfire{} arc is reliable or not. For this, we require a visually inspected `true' dataset to train upon. We presented a subsample of spiral galaxies from the \textit{stellar mass-limited sample} to volunteers in an interface which we called \spotter{}\footnote{www.zooniverse.org/projects/uon/spiral-spotter}, created using the \textsc{Zooniverse} project builder\footnote{www.zooniverse.org/lab}. Volunteers were presented with an image of a spiral galaxy, with the \sparcfire{} identified arcs overlaid. An example of one of these images is shown in Fig.~\ref{fig:spotter_subject}. In total, 252 galaxies were visually inspected by $\geq$3 volunteers, with 1617 spiral arcs visually inspected. The volunteers were asked about each \sparcfire{}-identified arc, with six possible responses. They could indicate that arcs were good matches to real spiral arms (good), poor matches to real arms (poor), matches to weak spiral arms (weak), extensions of previously identified arms (extension), fits to features that were not spiral arms (junk) or not present in the image (missing). The $N_\mathrm{clicks}$ column of Table~\ref{table:spotter} shows how the total number of votes were distributed for all arcs. Each arc is identified as one of the six categories, depending on which response recieved the greatest number of votes.  The number of arcs in each category is shown in column $N_\mathrm{arcs}$ of the table (any arcs where the majority vote was split between multiple categories were excluded). It is notable from this table that most of the arcs that \sparcfire{} identifies are not good matches to real spiral arcs: only 15\% of arcs were classified as `good'. It is therefore imperative that we identify a technique that removes the poorly matched spiral arcs in \sparcfire{}, which we discuss in the next section.

\subsubsection{Applying \spotter{} to the full dataset}
\label{sec:spotter_application}

From the \spotter{} statistics, we selected good arcs as those where the majority of volunteers indicated that they were good matches to true spiral arms in galaxy images. For each arc we thus have a label of whether it visually corresponds to a real spiral arm or not. We trained two models with the aim of selecting only the spiral arcs from \sparcfire{} that correspond to real arm features. The first model simply aimed to identify a suitable length cut to select only the longest arcs (as in \citealt{Davis_14}). The second model used a more sophisticated support vector machine approach (SVM) from the \texttt{scikit-klearn} package \texttt{SVM.SVC} \citep{sklearn} trained upon more of the properties associated with each arc. For a more complete description on how this method was trained, we refer readers to appendix~\ref{appendix:SVM}.

In order to assess how well a classifier is doing, we use two statistics, completeness and contamination. Our completeness is given by \begin{equation}
\label{eq:completeness}
\mathrm{completeness} = \frac{N (\mathrm{good \, arcs, \, \mathrm{classifier}})}{N (\mathrm{good \, arcs, \, \mathrm{inspected}})} , 
\end{equation} and our contamination is given by \begin{equation}
\label{eq:contamination}
\mathrm{contamination} = \frac{N (\mathrm{rejected \, arcs, \, \mathrm{inspected}})}{N (\mathrm{good \, arcs, \, \mathrm{classifier}})} . 
\end{equation} In theory, there are two ways in which a classifier can be tested -- either the completeness can be maximised or the contamination minimised. There is a trade-off between these statistics, in that including more positives in a sample will generally improve the completeness, but also increase the level of contamination. Given that we have a large number of galaxies in our samples to compare, ensuring a high level of completeness is not critical to this paper. Instead, we wish to ensure that any sample we do define is as clean as possible, so that any arc measurements are as reliable as possible. We therefore aim to classify \sparcfire{} arcs to decrease the level of contamination.

A useful piece of information that we can also use to identify real arc features is galaxy chirality (whether arcs wind clockwise or anti-clockwise), assuming that all of the arms in spiral galaxies can only wind clockwise or anti-clockwise. The \sparcfire{} suggested statistic that best agrees with the GZ1 measured chiralities is the `weighted pitch angle sum'. The sum of all pitch angles is calculated (with clockwise arcs given positive values and anticlockwise arcs given negative values), and weighted by the arc length. If the sum is positive, then the a galaxy is deemed to have clockwise dominant chirality, and if it is negative, the dominant chirality is deemed to be anticlockwise. From the galaxies in our \textit{stellar mass-complete sample}, 4801 were fit in \sparcfire{} (fit state=`OK'), of which 4779 (99.5\%) were visually classified by $\geq$5 people in Galaxy Zoo 1 (GZ1; \citealt{Lintott_11}). We see a strong agreement between the \sparcfire{} and GZ1 measured chiralities, with 4112/4779 (85.8\%) galaxies in agreement, or 3676/3967 (92.5\%) when considering only galaxies where $\geq$80\% of GZ1 classifiers agree. We therefore have the option to remove any arcs which do not agree with dominant chirality of the galaxy as measured by \sparcfire{}, if we wish to clean our sample. We note that there are rare cases where both chiralities exist in galaxies -- such galaxies are likely to be disturbed galaxies, which are not the main focus of this paper, and would require a more detailed examination.

\begin{figure*}
    \includegraphics[width=0.975\textwidth]{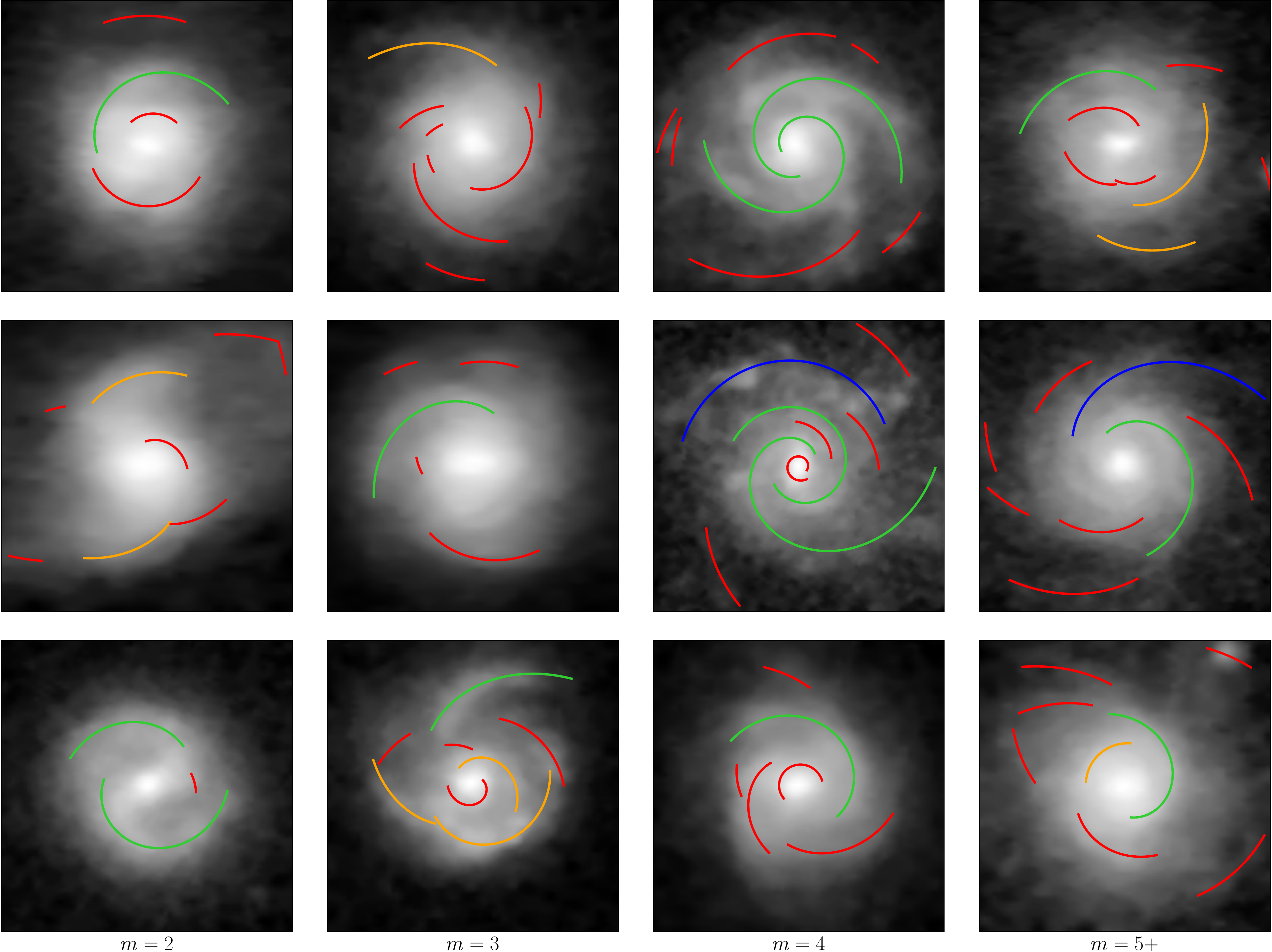}
    \caption{A randomly selected subsample of galaxies from the \spotter{} sample with $m$=2, 3, 4 or 5+ spiral arms as identified by GZ2. Arcs are coloured by their status as true positive (\spotter{}=good arc, SVM=good arc, lime green arcs), false positive (\spotter{}=poor arc, SVM=good arc, blue), false negative (\spotter{}=good arc, SVM=poor arc, orange) and true negative (\spotter{}=poor arc, SVM=poor arc, red).}
    \label{fig:arc_images}
\end{figure*}

Using a simple threshold to measure arcs, \citet{Davis_14} suggest that 75 pixels is the best length for finding a good agreement between \textit{arc number} and \textit{arm number} as measured by GZ2. Applying this threshold to the 252 galaxies in the \spotter{} subset achieves a completeness of 0.97 (0.92 only selecting arcs which agree with the dominant chirality) and a contamination of 0.57 (0.51). Using the trained SVM method, we achieve a completeness of 0.75 (0.73 only selecting the dominant chirality) and contamination of 0.19 (0.19). For comparison, a length cut of 125 pixels achieves a similar level of completeness of 0.74 (0.72), but suffers from a greater level of contamination, with values of 0.34 (0.28). 
Given the statistics listed above, the trained SVM method was preferred as it minimised the level of contamination better than the simple length cut. We also removed any arcs which did not agree with the dominant galaxy chirality, as we expect all arcs within a galaxy to have a single chirality. In total, 3028/6222 of the spiral galaxies from the \textit{stellar mass-complete sample} had one or more reliably identified spiral arc. For the entire sample of arcs in the \spotter{} set, there are 163 true positives (\spotter{}=good arc, SVM=good arc), 39 false positives (\spotter{}=poor arc, SVM=good arc), 83 false negatives (\spotter{}=good arc, SVM=poor arc) and 1332 true negatives (\spotter{}=poor arc, SVM=poor arc). Some examples of the \spotter{} galaxy images with their \sparcfire{} identified spiral arcs are shown in Fig.~\ref{fig:arc_images}.

\subsubsection{Checking for redshift dependent bias}
\label{sec:redshift_dependent_bias}

\begin{figure}
    \includegraphics[width=0.45\textwidth]{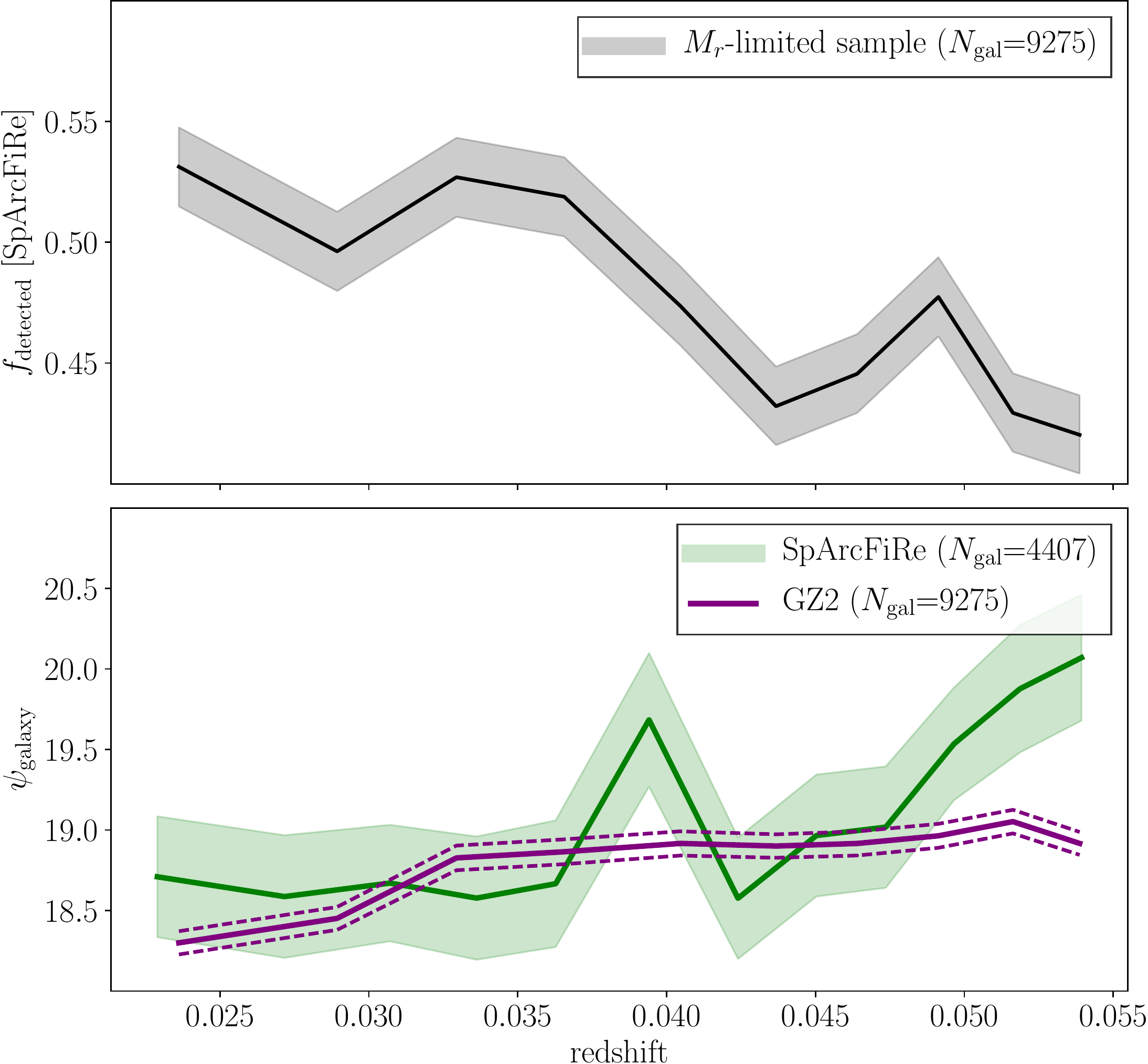}
    \caption{(a) Fraction of galaxies with at least one good arc as a function of redshift. The solid line indicates the fraction in each bin, and the shaded region shows the error as calculated from the method described in \citet{Cameron_11}. Fewer galaxies have detected arcs in \sparcfire{} at higher redshift. (b) Average pitch angle as a function of redshift for \sparcfire{} measured pitch angles (green line with filled errors) and GZ2 measured pitch angles discussed in Sec.~\ref{sec:sparcfire_vs_gz2} (purple line with dashed errors). The lines show the mean for each bin, and the errors indicate one standard error on the mean.}
    \label{fig:z_vs_fP}
\end{figure}

In Galaxy Zoo 2, the effects of redshift-dependent \textit{classification bias} are carefully considered. In \citealt{Hart_16} we developed a technique for modeling and removing biases due to resolution and signal-to-noise effects, building on the work of \citet{Bamford_09} and \citet{Willett_13}. This ensures that our GZ2 classifications are stable with redshift (see Fig.~8 of \citealt{Hart_16}). However, such biases are not unique to visual classifications. The fraction of galaxies for which SpArcFiRe finds at least one `reliably identified' arc (as defined in Section 2.3.2) is plotted as a function of redshift in Fig.~\ref{fig:z_vs_fP}a. We use a luminosity-limited sample for this analysis, complete for all galaxies in the redshift range $0.02 < z \leq 0.055$ brighter than $M_r=-19.95$. There are 9275 galaxies in this sample. We see that fewer galaxies have arcs detected by \sparcfire{} at higher redshift; $59 \pm 3$ per cent of our sample have one or more arcs in the lowest redshift bin at $z \sim 0.02$, whereas only $41 \pm 2$ per cent have detected arcs at the high redshift limit of the sample at $z \sim 0.055$. \sparcfire{} rarely detects all of the spiral arms in a galaxy, as can be seen in Fig.~\ref{fig:arc_images}, and is less likely to reliably detect spiral arms in lower resolution images, resulting in greater incompleteness at higher redshift. 

To address this issue, in Sec.~\ref{sec:sparcfire_vs_gz2} we develop an alternative method of determining spiral galaxy pitch angle using the GZ2 statistics alone, which can be applied to all the spiral galaxies in our sample. We present all our results using both measures of pitch angle, finding good agreement.

As a further check of whether redshift-dependent incompleteness or measurement issues could bias our results (e.g., if we are more likely to lose galaxies with looser or tighter arms), we examine the measured pitch angles as a function of redshift in Fig.~\ref{fig:z_vs_fP}b. Here we see that there is no significant redshift trend for either \sparcfire{}- or GZ2-derived pitch angles. These checks reassure us that our results are robust to the details of the pitch angle measurements.

\section{Results}
\label{sec:results}

In this section, we use the \sparcfire{} arc measurements detailed in Sec.~\ref{sec:identifying_arms} to identify the geometry of the spiral arms in our GZ2 derived spiral galaxies. In particular, these data can be used in conjunction with our GZ2 derived spiral morphological statistics, complementing our already measured spiral arm numbers in Sec.~\ref{sec:arm_number} to gain a more complete insight into the spiral structure in galaxies. Links between spiral arm pitch angle and other galaxy properties, including visual morphological characteristics from GZ2, mass properties from \citet{Mendel_14} and galaxy star formation rates (SFRs) are investigated in this section.

\subsection{Pitch angle distributions}
\label{sec:psi_distributions}

\begin{figure}
    \includegraphics[width=0.45\textwidth]{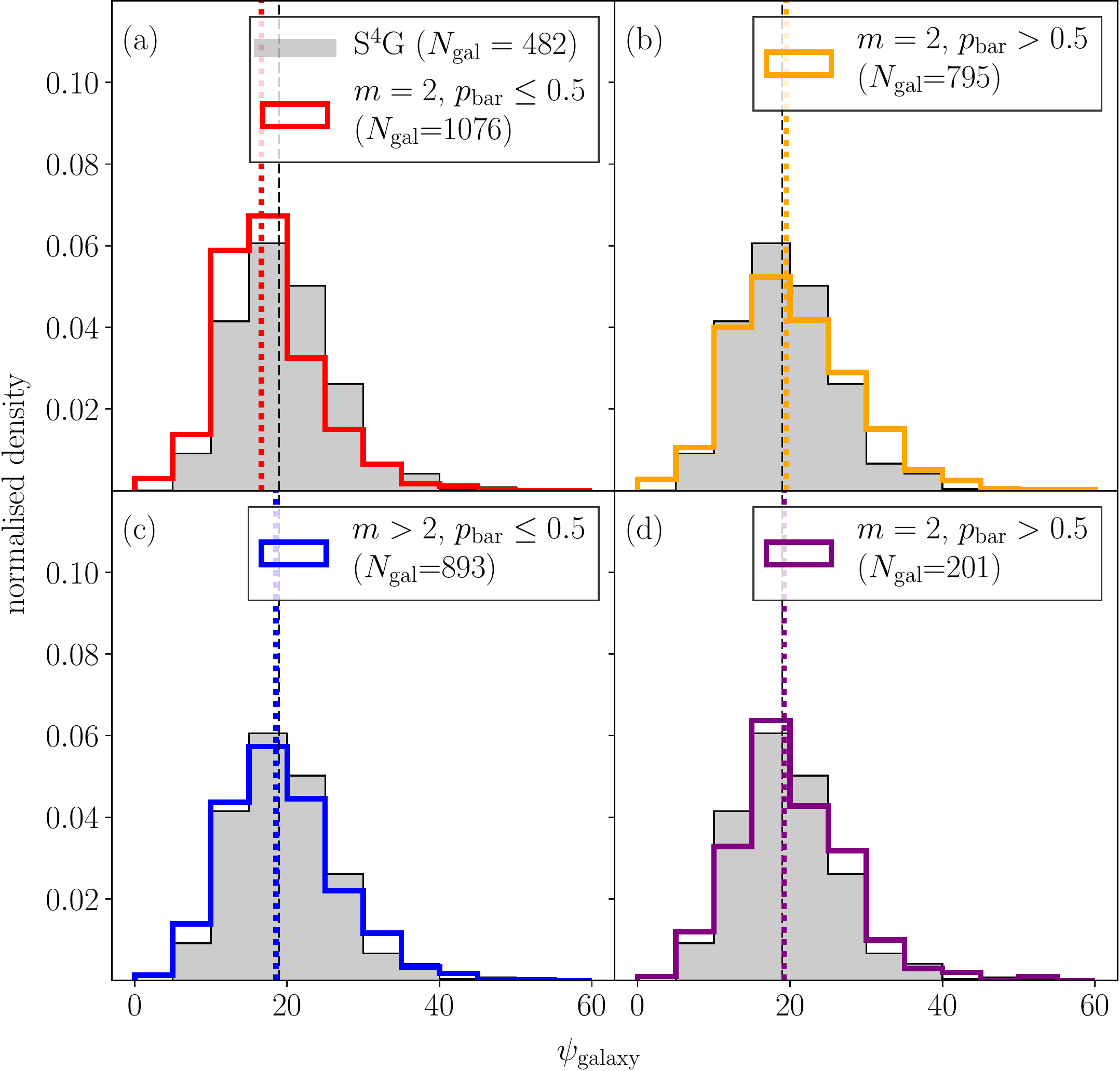}
    \caption{Distributions of \sparcfire{} derived galaxy pitch angles ($\psi_\mathrm{galaxy}$) for four samples of spiral galaxies: (a) two-arm weakly barred/unbarred, (b) two-arm barred, (c) many-arm weakly barred/unbarred and (d) many-arm barred. The grey histogram in each panel show the distributions for the S$^4$G sample of low-redshift galaxies. The vertical dashed black lines show the median pitch angle of the S$^4$G sample, and the dotted vertical coloured lines show the median for each of our subsamples.}
    \label{fig:P_histograms}
\end{figure}

Using the arcs identified in Sec.~\ref{sec:identifying_arms}, the overall pitch angles of our spiral galaxies are compared. We use two statistics to define pitch angles. The quantity $\psi_\mathrm{arc}$ is the pitch angle assigned to each arc. To define a galaxy-level pitch angle, $\psi_\mathrm{galaxy}$, we use the same length-weighted average pitch angle as \citet{Davis_14}, except we restrict the arcs to the ones we have defined as reliable. The statistic is defined by \begin{equation}
    \label{eq:pitch_angle}
    \psi_\mathrm{galaxy}  = \sum_{n=1}^{N_\mathrm{arcs}} \frac{L_n \psi_n}{L_\mathrm{total}} ,
\end{equation} where $N_\mathrm{arcs}$ is the total number of well-identified arcs, $L_n$ is the length of each individual arc, $\psi_n$ is the pitch angle of each detected arc and $L_\mathrm{total}$ is the sum of all of the arc lengths. In order to compare the distributions of pitch angles covering all of the broad morphological characteristics identified in GZ2, our \textit{stellar mass-complete} sample is divided into four categories: two-arm weakly barred/unbarred ($m=2$ and $p_\mathrm{bar}<0.5$), two-arm barred ($m=2$ and $p_\mathrm{bar} \geq 0.5$), many-arm weakly barred/unbarred ($m>2$ and $p_\mathrm{bar} < 0.5$) and many-arm barred ($m>2$ and $p_\mathrm{bar} \geq 0.5$), and the distributions of $\psi_\mathrm{galaxy}$ are shown in Fig.~\ref{fig:P_histograms}. Our mean pitch angle is 18.0\textdegree{} with 16th and 84th percentiles of 12.2\textdegree{} and 26.1\textdegree{} for the entire sample of spirals. For comparison, the \citet{Herrera_Endoqui_15} S$^4$G sample is shown. In this case, we have no arm lengths, so we measure $\psi_\mathrm{galaxy}$ as the mean pitch angle of all of the arcs in each galaxy. For this comparison sample, the mean pitch angle is 19.0\textdegree{}, with 16th and 84th percentiles of 13.5\textdegree{} and 25.7\textdegree{}. We see that the overall distributions match well with observed spiral arms in S$^4$G, with the peak pitch angles at $\sim$ 15-20\textdegree{} in all cases and KS $p$-value > $10^{-2}$ in all but the two-arm unbarred subsample, where the distribution is clearly offset to smaller pitch angles. We note that we also see very few galaxies with $\psi_\mathrm{galaxy}<$10\textdegree{} and $\psi_\mathrm{galaxy}>$40\textdegree{}, as expected from observations of nearby galaxies \citep{Seigar_08}. 

\subsection{Comparing \sparcfire{} and GZ2 derived pitch angles}
\label{sec:sparcfire_vs_gz2}

In order to check the reliability of our pitch angle measurements, we wish to compare our pitch angles to independently derived pitch angle measurements. In GZ2, there are two characteristics of the spiral structure that have been classified by eye: the number of spiral arms and how tightly wound the spiral arms are. The latter gives a qualitative measure of pitch angle in galaxies. In GZ2, volunteers were asked whether the arms they saw were `tight', `medium' or `loose'. Here, we expect that galaxies classified with `loose' arms to have larger galaxy pitch angles.

\begin{figure}
    \includegraphics[width=0.45\textwidth]{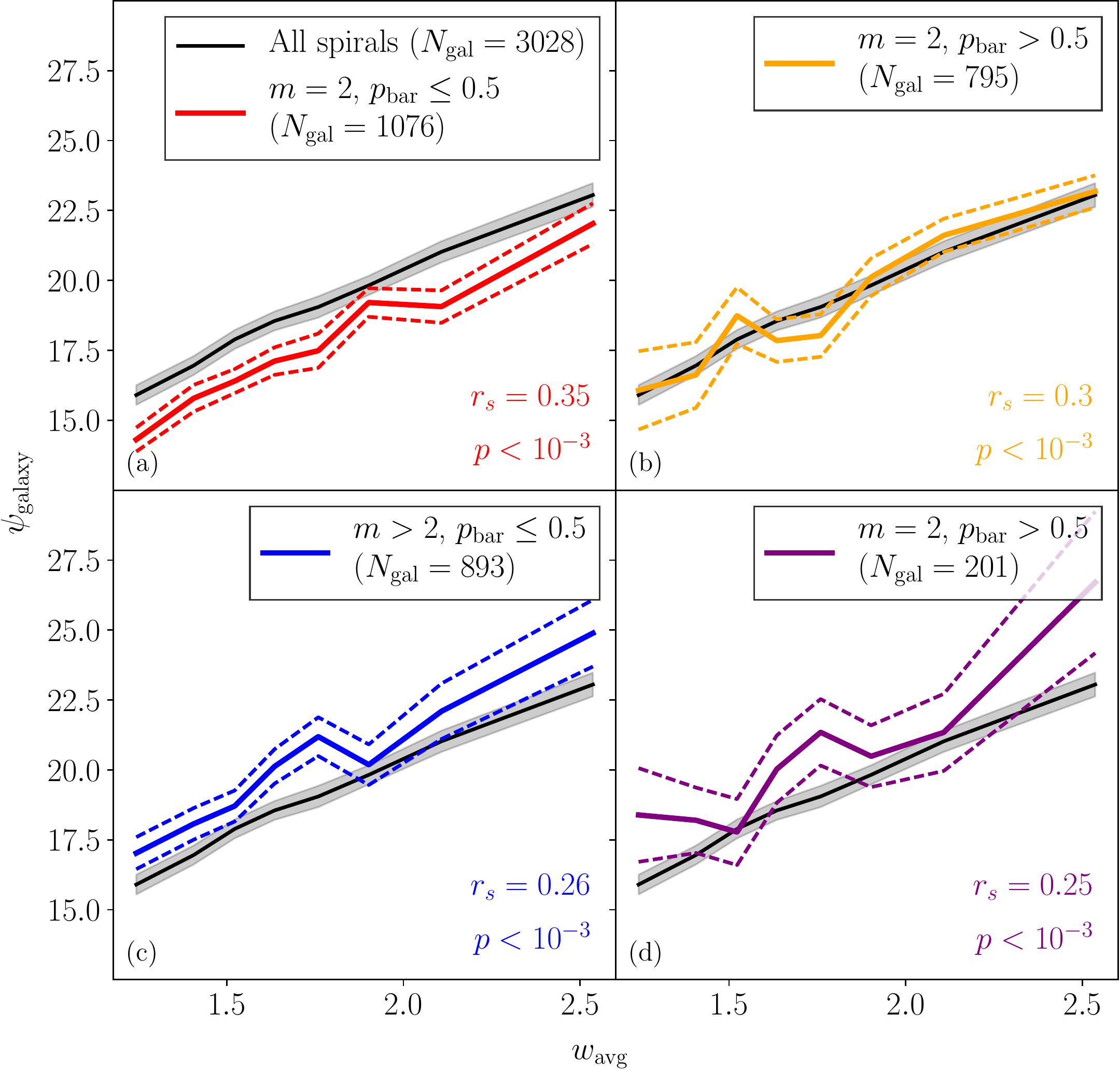}
    \caption{GZ2 measured arm tightness ($w_\mathrm{avg}$) vs. \sparcfire{} measured galaxy pitch angle ($\psi_\mathrm{galaxy}$) for the \textit{stellar mass-complete sample} of spirals. The lines indicate the mean value for each bin, and the errors indicate 1 standard error on the mean. The black line with grey-filled errors represent the full \textit{stellar mass-limited spiral sample}, and the thicker coloured lines with dashed errors show the same values for four subsamples (a) two-arm weakly barred/unbarred, (b) two-arm barred, (c) many-arm weakly barred/unbarred and (d) many-arm barred. A strong correlation is observed between the GZ2 arm winding statistic and the measured pitch angle in all cases.}
    \label{fig:w_avg_vs_P}
\end{figure}

To measure how tightly wound the spiral structure is in GZ2, we define two statistics. The first is $w$, which is defined as the response to the arm winding question which had the highest debiased vote fraction, and can take the values `t', `m' or `l' (tight, medium or loose). The second statistic we define is the average winding score, $w_\mathrm{avg}$. This is defined as \begin{equation}
    \label{eq:w_avg}
    w_\mathrm{avg} = \sum_{w=1}^{3} w p_w.
\end{equation} This statistic is analogous to the one defined in Sec.~\ref{sec:arm_number}, this time using the responses to the arm winding GZ2 question rather than the arm number question. If a galaxy has perfect agreement and all classifiers said the spiral arms were tightly wound, $w_\mathrm{avg}$=1, and if everyone classified the arms as loose, then $w_\mathrm{avg}=3$.
In Fig.~\ref{fig:w_avg_vs_P}, we compare the winding scores from GZ2 with the directly measured pitch angles, $\psi_\mathrm{galaxy}$, derived from \sparcfire{}. The black lines in each of the panels of Fig.~\ref{fig:w_avg_vs_P} represent the entire population of 3190 galaxies from the \textit{stellar mass-complete sample} with reliable arcs identified by \sparcfire{}, with no cuts made in arm number or bar probability. We see that a clear correlation does exist between the two statistics (Spearman rank statistics $r_s$=0.30, $p<10^{-3}$). Such a result is expected if both measurements are reliable methods for measuring spiral arm pitch angle. To check whether this relation holds for all types of spiral structure, we subdivide this full sample into the same four broad spiral morphological subsamples as in Sec.~\ref{sec:psi_distributions}. The winding score vs. pitch angle relation is plotted for each of these subsamples in Fig.~\ref{fig:w_avg_vs_P}. Here we see that the correlation between these two measures still exist ($r_s$=0.35, 0.30, 0.26 and 0.25, $p<10^{-3}$), no matter which type of spiral structure is present in the galaxy disc. These results therefore offer encouragement that the \sparcfire{} derived pitch angles are physically meaningful. It is also interesting to note that pitch angle estimates are also obtainable from the GZ2 data alone, given the tight relationship between the GZ2 and \sparcfire{} measured statistics. One can do this using a fit to the GZ2 data. A linear best fit line yields
\begin{equation}
\label{eq:predict_psi}
\psi_\mathrm{GZ2} = 6.37w_\mathrm{avg} + 1.30m_\mathrm{avg} + 4.34.
\end{equation} This calibration depends on both the GZ2 arm winding score and the arm number. From Fig.~\ref{fig:w_avg_vs_P}a and ~\ref{fig:w_avg_vs_P}c, including $m_\mathrm{avg}$ and $p_\mathrm{bar}$ we see an offset from the black line for all galaxies that depends on spiral arm number -- arm number is included in the fit to avoid a systematic uncertainty with arm number. From the distributions of $\psi_\mathrm{galaxy}$ vs. $\psi_\mathrm{GZ2}$, we find that the rms scatter between the two galaxy pitch angle measurements is $\pm7$\textdegree{}. Given that this covers a significant range of true observed pitch angles (see Fig.~\ref{fig:P_histograms}), we advise that these pitch angle measurements should not be used for small samples of galaxies. However, using Eq.~\ref{eq:predict_psi} on large samples of galaxies should give accurate measurements of galaxy pitch angle across the population, which can be seen from the tightness of the standard error on the mean in the black lines in Fig.~\ref{fig:w_avg_vs_P}.

\subsection{Pitch angle vs. galaxy structural parameters}
\label{sec:results_psi_vs_structure}

In this section, galaxy structural properties and their relation to spiral arm pitch angles are investigated. Of particular note are two statistics that have been derived from the GZ2 classifications of our galaxies: the number of spiral arms and the presence of bars in galaxy discs. 

\subsubsection{Spiral arm number}
\label{sec:results_arm_number}

Spiral arms can be categorised by their arm number. It is often suggested that there are a multitude of mechanisms that can lead to spiral patterns emerging in discs, and that the mechanisms responsible for grand design patterns differ from those that lead to many-armed ones (\citealt{Dobbs_14} and references therein). Simulations of modal spiral arms also predict many-arm structures will have looser spiral arms than spirals with fewer arms \citep{Grand_13}. Given that some two-arm structures are associated with galaxy-galaxy interactions, which are in turn associated with loose structures \citep{Casteels_13}, the two-arm population may include galaxies with looser arms. In this section, we compare the pitch angles for galaxies with different spiral arm numbers measured from the GZ2 statistics outlined in Sec.~\ref{sec:arm_number}.

\begin{figure*}
    \includegraphics[width=0.975\textwidth]{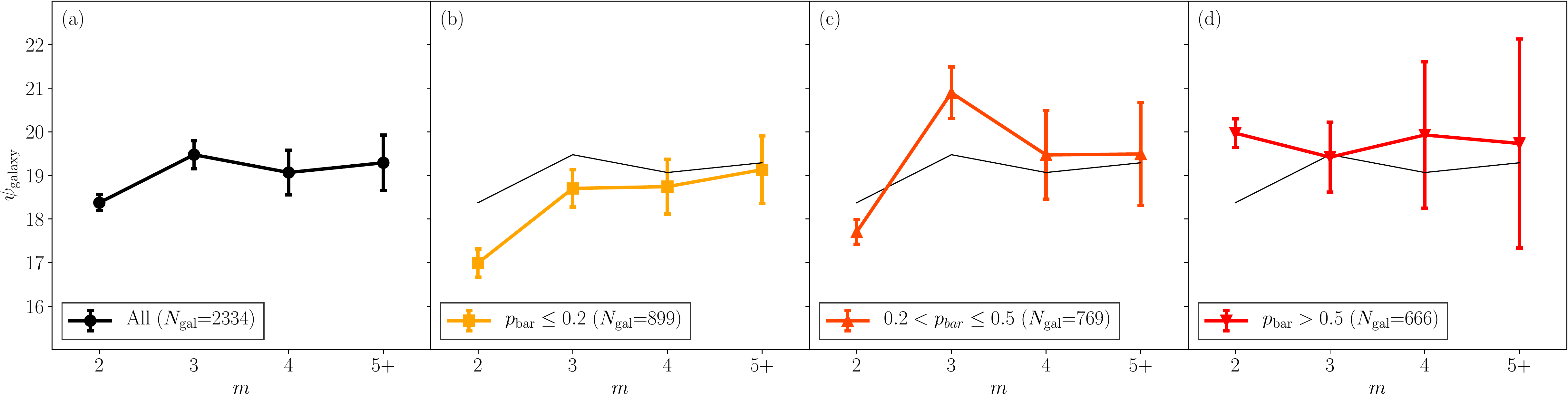}
    \caption{Spiral arm number, $m$, vs. pitch angle, $\psi_\mathrm{galaxy}$ for the \textit{stellar mass-matched spiral sample} with reliable \sparcfire{} arcs for (a) all, (b) unbarred, (c) weakly barred and (d) strongly barred galaxies. The coloured markers show the mean for each arm number, and the errorbars indicate one standard error on the mean. The black line in each plot shows the relationship for all spiral galaxies, irrespective of bar presence for reference. Two-arm galaxies have tighter spiral arms than many-armed galaxies for unbarred and weakly barred galaxies.}
    \label{fig:m_vs_P}
\end{figure*}

\begin{figure*}
    \includegraphics[width=0.975\textwidth]{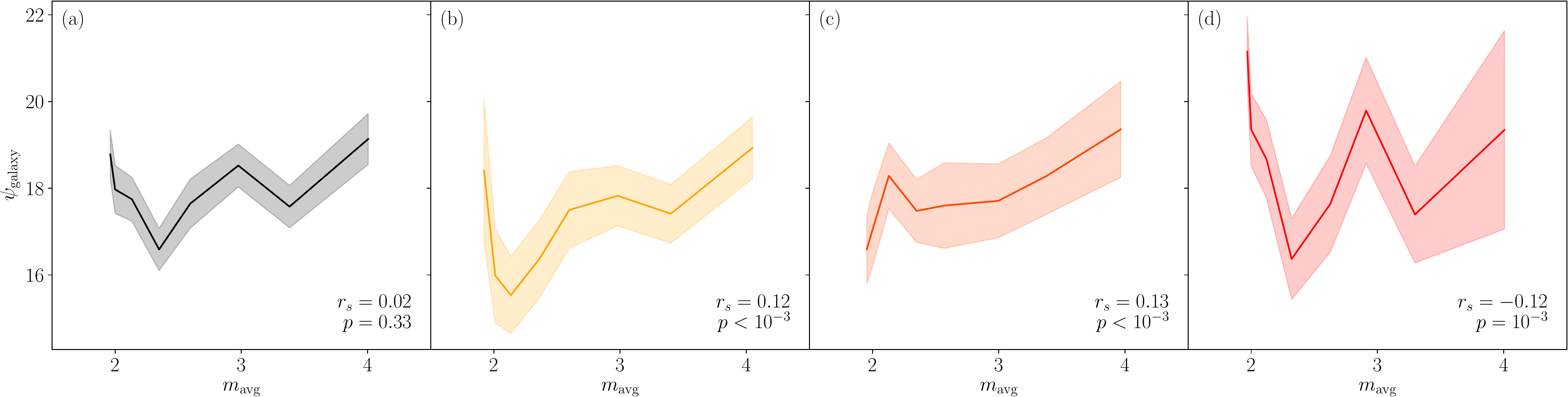}
    \caption{Weighted average arm number, $m_\mathrm{avg}$ vs. pitch angle, $\psi$, for the \textit{stellar mass-matched spiral sample} with reliable \sparcfire{} arcs for (a) all, (b) unbarred, (c) weakly barred and (d) strongly barred galaxies. The data is divided into eight bins in each panel, and the line shows the mean for each bin. The shaded error region shows one standard error on the mean. The arms of fewer-armed galaxies are tighter for both the unbarred and weakly barred subsamples.}
    \label{fig:m_avg_vs_P}
\end{figure*}

In Fig.~\ref{fig:m_vs_P}a, we plot spiral arm number, $m$, vs. pitch angle for all of our spiral galaxies. We use the \textit{stellar mass-matched sample} to do this, as galaxy stellar mass properties influence the pitch angles of spiral galaxies (e.g. \citealt{Seigar_06,Seigar_14}). Spiral galaxies with one spiral arm are removed from this analysis, as there are only 31 one-arm spirals with reliable arms in the \textit{stellar mass-matched sample}. Here we see a weak trend that galaxies with more spiral arms have looser spiral structures -- two-arm spirals have mean pitch angle $18.6\pm 0.2$\textdegree{}, whereas the corresponding values for each of the many-arm categories are $19.2\pm 0.3$, $19.2\pm 0.5$, $19.4\pm 0.6$\textdegree{} for $m$=3, 4 and 5+ respectively.

Bars could potentially influence the pitch angles of spiral arms, and are more common in grand design, two-arm spiral galaxies (e.g. \citealt{Elmegreen_82,Elmegreen_11}). We investigate the role of bars in more detail in Sec.~\ref{sec:results_bars}. In this section, we control for the bar influence on our arm number comparisons by using cuts on the GZ2 $p_\mathrm{bar}$ statistic described in Sec.~\ref{sec:bar_presence}. In Fig.~\ref{fig:m_vs_P}b-d, we show the arm number vs. pitch angle relationship for unbarred, weakly barred and strongly barred galaxies separately. Removing barred galaxies has little effect on the spiral arm pitch angle of many-arm galaxies: for three arm galaxies, the mean pitch angles are $18.7 \pm 0.4$\textdegree{}, $20.9 \pm 0.6$\textdegree{} and $19.4 \pm 0.8$\textdegree{} for unbarred, weakly-barred and strongly-barred galaxies. For four-armed galaxies, the mean pitch angles are $18.7 \pm 0.6$\textdegree{}, $19.5 \pm 1.0$\textdegree{} and $19.9 \pm 1.7$\textdegree{}, and for five or more armed galaxies they are $19.1 \pm 0.8$\textdegree{}, $19.5 \pm 1.2$\textdegree{} and $19.7 \pm 2.4$\textdegree{} respectively. However, we see that the galaxy pitch angle does depend on bar strength in two-arm galaxies: the mean pitch angles are $17.0 \pm 0.3$\textdegree{}, $17.7 \pm 0.3$\textdegree{} and $20.0 \pm 0.3$\textdegree{}. From Fig.~\ref{fig:m_vs_P}, we can see that two-arm galaxies are between 1.7\textdegree{} and 2.1\textdegree{} tighter than each of the many-arm subsamples.

In Fig.~\ref{fig:m_avg_vs_P}a-d, the spiral arm number vs. pitch angle relation is investigated, this time using the average arm number $m_\mathrm{avg}$, rather than the absolute arm number. Similar results are observed, where galaxies with more spiral arms have a tendency to have looser arms. As was the case in Fig.~\ref{fig:m_vs_P}a, a weak correlation is observed when we include all spiral galaxies in Fig.~\ref{fig:m_avg_vs_P}a ($r_s$=0.02, $p$=0.33). However, a clear trend is observed where the arms of many-arm spirals are looser than in two-arm spirals for unbarred ($r_s$=0.12, $p<10^{-3}$)and weakly barred galaxies ($r_s$=0.13, $p<10^{-3}$) in Fig.~\ref{fig:m_avg_vs_P}b-c. The trend disappears when one considers strongly barred galaxies and galaxies with fewer arms actually have looser pitch angles ($r_s=-0.12$, $p=10^{-3}$).

\subsubsection{The influence of bars}
\label{sec:results_bars}

\begin{figure}
    \includegraphics[width=0.45\textwidth]{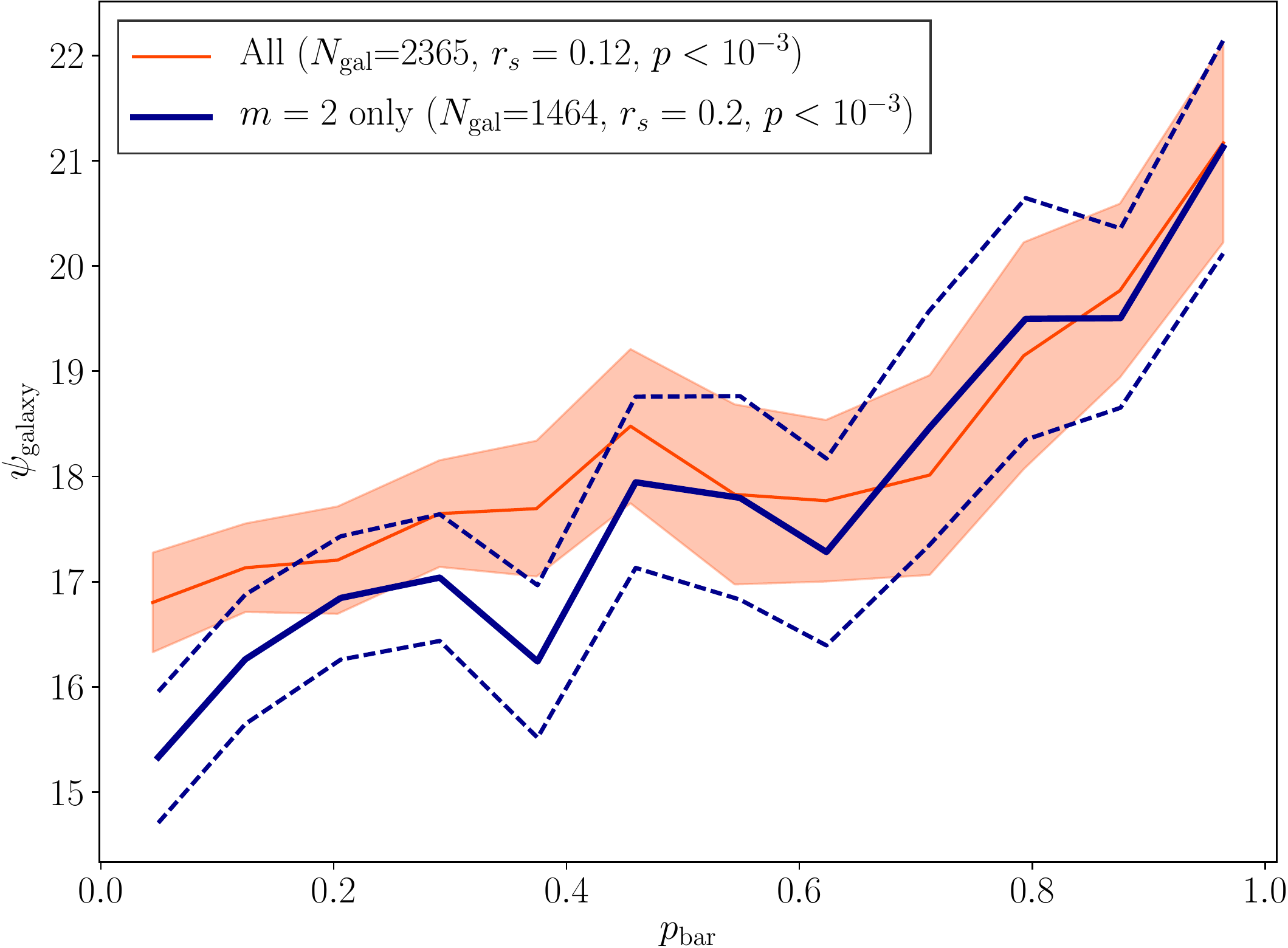}
    \caption{GZ2 bar fraction, $p_\mathrm{bar}$ vs. spiral arm pitch angle, $\psi_\mathrm{galaxy}$. The orange line with filled errors indicates the relation for all spiral galaxies in the \textit{stellar mass-limited sample}. The thicker blue line with dashed errors show the same relation for only galaxies with two spiral arms in GZ2 ($m=2$). The lines indicate the mean for each of the bins, and the errors show one standard error on the mean. Galaxies with stronger bars have looser spiral arms.}
    \label{fig:p_bar_vs_P}
\end{figure}

Bars can affect the types of spiral structures observed in galaxies. In Fig.~\ref{fig:p_bar_vs_P}, we plot the GZ2 measured bar fraction, $p_\mathrm{bar}$ for two subsamples of galaxies with measured pitch angles from \sparcfire{}. The thin orange line with filled errors shows how the galaxy pitch angle depends on the GZ2 bar probability for all galaxies in the \textit{stellar mass-complete spiral sample}, irrespective of spiral arm number. Here we observe a correlation, where galaxies with stronger bars tend to have looser arms ($r_s=0.12$). If we instead focus on only the galaxies with two spiral arms, indicated by the thicker red line in Fig.~\ref{fig:p_bar_vs_P}, then a stronger correlation emerges ($r_s=0.20$). For all spiral galaxies, the mean pitch angle varies from $17.9 \pm 0.4$\textdegree{} for the bin with the lowest bar fraction to $22.1 \pm 0.7$\textdegree{} for the bin with the highest bar fraction, a difference of 4.1\textdegree{}. In the two-arm case, it varies between $16.2 \pm 0.5$\textdegree{} and $22.1 \pm 0.8$\textdegree{}, a significant difference of 5.9\textdegree{}.

Here, we see two competing effects which affect the galaxy pitch angle. In Sec.~\ref{sec:results_arm_number}, a difference with respect to arm number was only observed in weakly barred or unbarred galaxies. Although two-arm spirals generally have tighter pitch angles, bars are also more common in these galaxies. When considering unbarred spirals, we have a population of two-arm spirals with arms with tight pitch angles, and a many-arm population with arms with looser pitch angles. Adding barred galaxies introduces a population of galaxies with looser arms, which preferentially have two spiral arms. This means the two-arm population has only slightly tighter spiral arms when one considers the overall population including barred galaxies.

\subsection{Galaxy stellar mass properties}
\label{sec:results_mass}

There is evidence that the central mass concentration affects the shear in galaxy discs, which in turn directly influences the spiral arm pitch angle, both in grand design spirals \citep{Seigar_06,Seigar_14} and in modal many-arm structures \citep{Grand_13}. Using the stellar mass properties of galaxies from \citet{Mendel_14}, we now investigate any correlations between central mass concentration and spiral arm structure.

\begin{figure*}
    \includegraphics[width=0.975\textwidth]{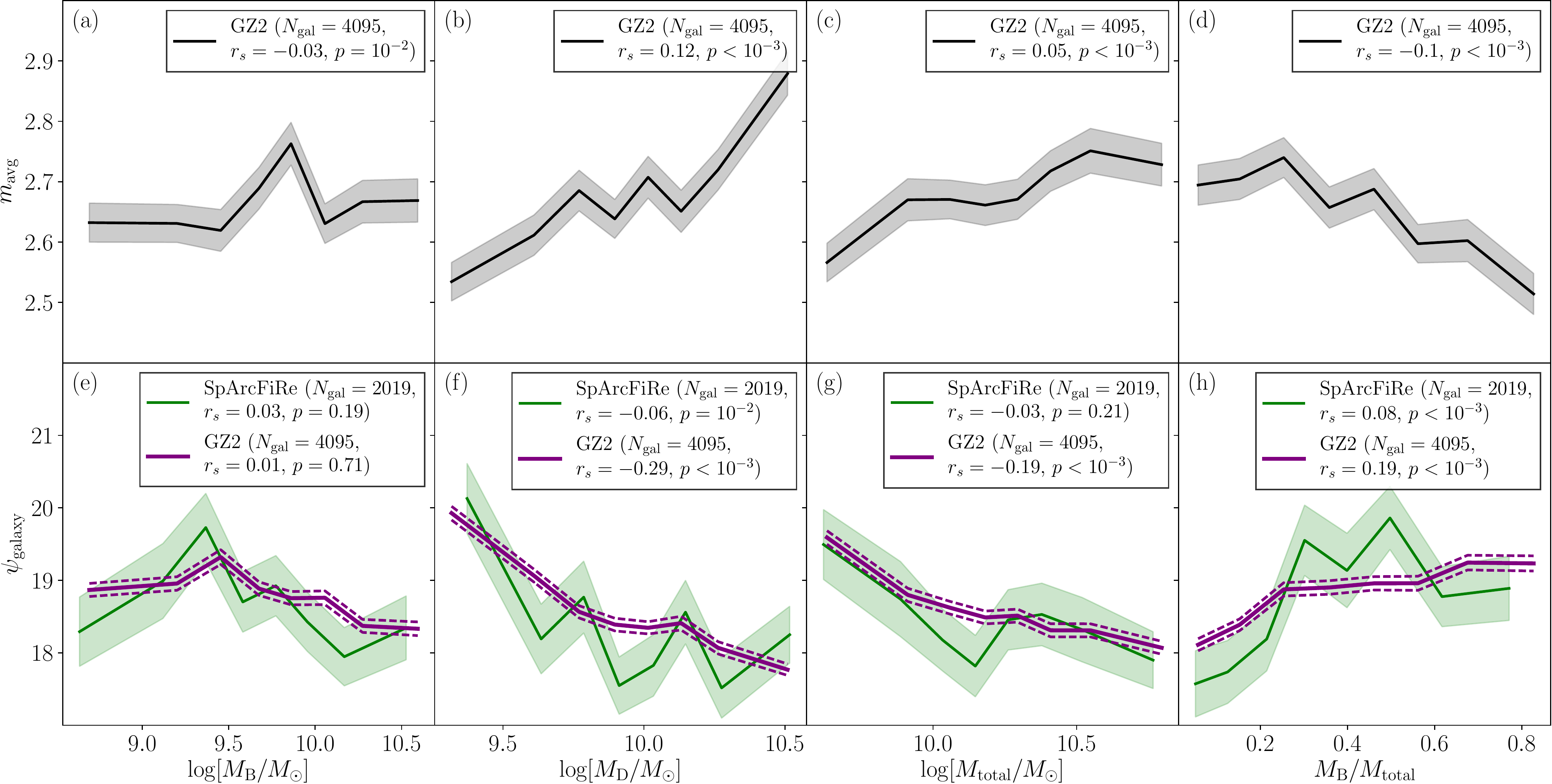}
    \caption{(a)-(d) Galaxy mass parameters vs. average arm number ($m_\mathrm{avg}$) for the \textit{stellar mass-complete sample}. Left to right: galaxy bulge mass, galaxy disc mass, galaxy total mass, galaxy bulge fraction. The black line shows the mean and the shaded black region indicates one standard error on the mean. Each galaxy is weighted by $1/V_\mathrm{max}$. (e)-(h) Galaxy mass parameters vs. pitch angle for the \textit{stellar mass-complete sample}. The thinner green line with shaded errors show the pitch angles derived from \sparcfire{}, and the thicker purple line with dashed errors show pitch angles measured from the GZ2 calibration. The strongest trends we observe are that galaxies with more massive discs have more spiral arms and tighter pitch angles.}
    \label{fig:bulge_disc}
\end{figure*}

Some of the spiral galaxies in our sample include bars, which often require a separate component to be fit \citep{Gadotti_09,Kruk_17}, potentially affecting the accuracy of bulge and disc mass measurements. For this reason, any strongly barred galaxies are removed from this analysis (removing galaxies with $p_\mathrm{bar} \geq 0.5$). Since all of our galaxies are visually classified spirals, then it is expected that they should be bulge-disc systems, with two distinct components. We therefore use all galaxies in the \textit{stellar mass-complete sample} which have a bulge and disc component fit. This leaves us with a sample of 4095 unbarred spirals, 2019 of which have spiral arm measurements in \sparcfire{}. The spiral arm number is compared for four mass characteristics of our \textit{stellar mass-complete} spiral sample: the bulge mass ($\log[M_\mathrm{B}]$), disc mass ($\log[M_\mathrm{D}]$), the total mass ($\log[M_\mathrm{total}]=\log[M_\mathrm{B}+M_\mathrm{D}]$) and the bulge-to-total ratio ($M_\mathrm{B}/M_\mathrm{total}$). These quantities are compared to the average arm number in Fig.~\ref{fig:bulge_disc}a-d. A clear trend is observed in Fig.~\ref{fig:bulge_disc}, where galaxy mass concentration does seem to have a connection to spiral arm number: galaxies with greater bulge fractions tend to have fewer spiral arms ($r_s=-0.10$, $p<10^{-3}$). We also see a weak trend that more massive galaxies also tend to have more spiral arms ($r_s=0.05$, $p=10^{-3}$), a result we first observed in \citet{Hart_16}, albeit with a different stellar mass indicator. From Fig.~\ref{fig:bulge_disc}a-b, we can see that these results appear to be driven by differences in disc mass, rather than bulge mass -- we see a stronger positive trend that galaxies with more massive discs have more spiral arms ($r_s=0.12$, $p<10^{-3}$) and a much weaker trend that galaxies with less massive bulges have fewer spiral arms ($r_s$=-0.03, $p=10^{-2}$). Galaxy mass properties do seem to affect the spiral arm number of galaxies but these differences are mainly due to galaxy disc mass variations, rather than any variations in bulge mass.

In Fig.~\ref{fig:bulge_disc}e-h, the same four mass characteristics are plotted against spiral arm pitch angle. We have two measurements of spiral arm pitch angle. The first, $\psi_\mathrm{galaxy}$ from \sparcfire{} is a directly measured quantity, but is only available for the 2019 galaxies with measured good arms in \sparcfire{}. The alternative GZ2 derived pitch angle (see Sec.~\ref{sec:sparcfire_vs_gz2}) is available for all 4095 spiral galaxies. As was the case with respect to spiral arm number, we do see correlations with galaxy mass. Here we see a weak positive trend that galaxies with greater bulge fractions tend to have looser spiral arms ($r_s$=0.08 for \sparcfire{}, 0.19 for GZ2; $p<10^{-3}$ in both cases). We see weak negative correlations that galaxies with greater total stellar mass have tighter arms ($r_s$=-0.03 and -0.19, $p=0.21$ and $<10^{-3}$). Comparing Fig.~\ref{fig:bulge_disc}e-f shows that it is the galaxy disc mass that is responsible for these trends, as was the case for the dependence of spiral arm number on stellar mass. There is a negative correlation between disc mass and pitch angle ($r_s$=-0.06 and -0.29, $p=10^{-2}$, $<10^{-3}$), but there is little or no correlation between bulge mass and galaxy pitch angle ($r_s$=0.03 and 0.01, $p=$0.19 and 0.71). It is interesting to note that these trends are the opposite to what we expect if pitch angle differences were driven purely by spiral arm number. In Fig.~\ref{fig:m_vs_P} and Fig.~\ref{fig:m_avg_vs_P} we saw that galaxies with more spiral arms had looser structures, so one would expect galaxies with more massive discs to have looser spiral arms. Galaxy central mass concentration does seem to affect both arm number and pitch angle, but the disc mass is the main reason for the observed differences, rather than the bulge mass.

\subsection{Star formation rates}
\label{sec:SFRs}

\begin{figure*}
    \includegraphics[width=0.975\textwidth]{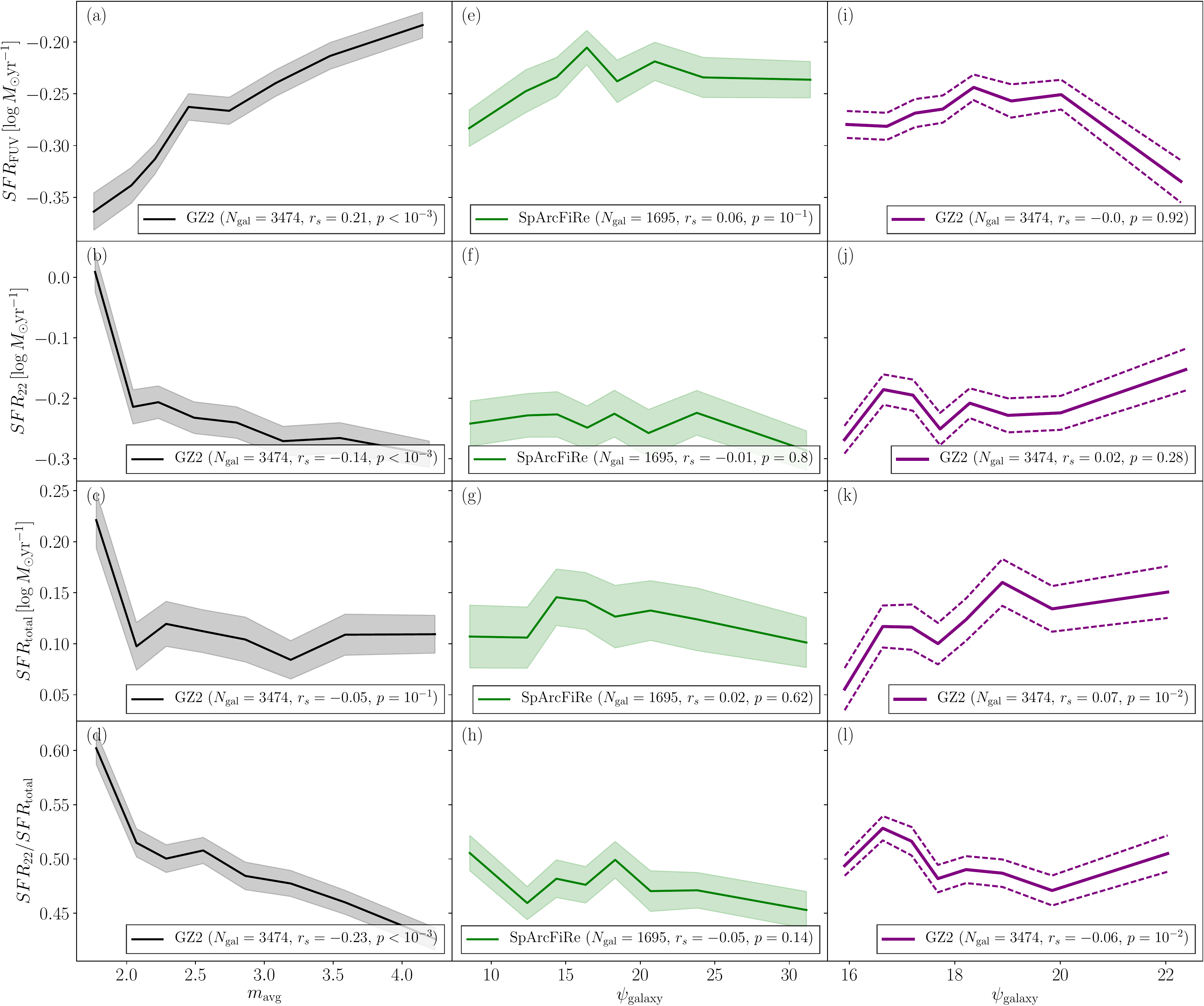}
    \caption{(a)-(d) Spiral arm number vs. galaxy SFR properties for the \textit{stellar mass-matched sample} of spirals. From top to bottom, the properties are far-UV SFR, mid-Ir SFR, total SFR and mid-IR SFR fraction. The same four quantities are plotted vs. \sparcfire{} derived pitch angles in (e)-(h) and GZ2 derived pitch angles in (i)-(l). The solid line in each subplot shows the mean, and the errors indicate one standard error on the mean. Although SFR properties vary with spiral arm number, there is no clear correlation with pitch angle.}
    \label{fig:sfrs}
\end{figure*}

The star formation properties of galaxies have been shown to correlate to the properties of the spiral arms. For example, in \citet{Hart_17}, we showed that galaxies with fewer spiral arms have a greater level of dust obscured star formation. Studies of grand design spirals also show that SFR is lower in galaxies with high shear \citep{Seigar_05}, which in turn means that spirals with the tightest arms have lower SFRs. The star formation properties of spiral galaxies are compared with respect to both the spiral arm number and the spiral arm pitch angle in galaxies. 

We use two photometrically derived SFRs in this paper. The first is the far-UV measured SFR, $SFR_\mathrm{FUV}$, which measures the amount of unobscured star formation in galaxies. The second is the mid-IR measured SFR, $SFR_\mathrm{22}$, which measures the amount of dust obscured star formation. These measures can be added to measure the total SFR, $SFR_\mathrm{total}$, and the ratios of these indicators describe the fraction of the star formation that is dust obscured. In the left hand side of Fig.~\ref{fig:sfrs}, four measures of SFR are presented as a function of weighted average arm number, $m_\mathrm{avg}$. They are $SFR_\mathrm{FUV}$, $SFR_\mathrm{22}$, $SFR_\mathrm{total}$, and the fraction of the SFR which is obscured ($SFR_\mathrm{22}/SFR_\mathrm{total}$) vs. average arm number. We use the \textit{stellar mass-matched sample} of $m$=1, 2, 3, 4 and 5+ arm galaxies for this analysis, as there is a stellar mass dependence on total SFR  which we wish to control for (e.g. \citealt{Brinchmann_04,Salim_07}). Fig.~\ref{fig:sfrs}a-d show the already established correlation seen in \citet{Hart_17}, this time using a continuous statistic to measure arm number. Galaxies with fewer spiral arms have more mid-IR star formation ($r_s$=-0.14, $p<10^{-3}$) and less far-UV star formation ($r_s$=-0.21, $p<10^{-3}$). The total SFRs remain consistent, with a sharp upturn below $m_\mathrm{avg}$=2. One-arm spirals are associated with galaxy-galaxy interactions \citep{Casteels_13}, which explains this trend. 

In  Fig.~\ref{fig:sfrs}e-l, we plot the same four SFR quantities vs. spiral arm pitch angle, rather than arm number. We again use two measures of spiral arm pitch angle:  Fig.~\ref{fig:sfrs}e-h show the \sparcfire{} derived pitch angles Fig.~\ref{fig:sfrs}i-l show the GZ2 derived pitch angles. Here we see no strong correlations that SFR varies in galaxies with different pitch angles, measured from both GZ2 and \sparcfire{}. Although spiral arm number has a strong influence on the amount of dust obscured star formation in galaxies, galaxies with different pitch angles all show consistent SFRs and fractions of obscured star formation.


\section{Discussion}
\label{sec:discussion}

In this paper, the \sparcfire{} method for measuring spiral arm pitch angles was used in conjunction with morphological statistics from GZ2. After evaluating the best possible method for measuring spiral pitch angles, the overall properties of galaxies were compared with respect to their spiral arm pitch angles. Galaxies with more spiral arms tend to have looser arms (larger pitch angle) -- arms associated with two arm structures are $\approx$2\textdegree{} tighter than those in many-armed galaxies. Galaxies hosting bars also have looser spiral arms: there is a positive correlation between the GZ2 $p_\mathrm{bar}$ statistic and pitch angle. Galaxies with the strongest bars have pitch angles $\approx$ 4\textdegree{} larger for all spirals, and $\approx$6\textdegree{} larger in two-arm spirals. Comparing the overall structures of galaxies, we find that galaxies with more massive discs have more spiral arms and smaller pitch angles.

\subsection{\sparcfire{} derived spiral arms}
\label{sec:sparcfire_derived_arms}

In Sec.~\ref{sec:psi_distributions}, the overall spiral arm pitch angle distributions were compared to those from \citet{Herrera_Endoqui_15}. Using the raw \sparcfire{} output combined with our by-eye classified galaxies, we are able to see that the range of spiral arms that our SVM deems reasonable agrees well with another survey of spiral arm pitch angle, S$^4$G \citep{Herrera_Endoqui_15}. Given that S$^4$G \citep{Sheth_10} is a volume-limited local survey with no cuts on galaxy morphology and \citet{Herrera_Endoqui_15} fitted their sample of 482 spirals by eye, this should provide a set of reliable arc measurements for a representative sample of local spirals. This means that the spiral arms being detected by \sparcfire{} do seem to have the same range of pitch angles as those observed using other methods. We see that there are very few galaxies with $\psi_\mathrm{galaxy}<10$\textdegree{} (7.5\% of galaxies) or $\psi_\mathrm{galaxy}>40$\textdegree{} (1.1\% of galaxies). This agrees remarkably well with both the most open spiral structure of 40-45\textdegree{} \citep{Seigar_98b,Block_99,Seigar_05}, and the tightest of 7-10\textdegree{} \citep{Block_99,Seigar_05} observed in other studies of nearby spirals. \citet{Seigar_08} attributed these limits to the limits of the shear in spiral galaxies -- the shear in the discs of galaxies is closely related to the central mass concentration and galaxies with little or no bulge have $\psi \approx 10$\textdegree. Conversely, the tightest observed pitch angles correspond to the largest central mass concentrations observed in galaxies. Our distributions of spiral arm pitch angles imply that these naturally occurring limits do exist in a statistically complete sample of spiral galaxies in the local Universe, and  that these limits extend to all types of spiral structure, rather than just grand design spirals. 

\subsection{Pitch angle and galaxy structure}
\label{sec:pitch_angle_vs_structure}

\subsubsection{Spiral arm number}
\label{sec:discussion_arms}

In Sec.~\ref{sec:results_psi_vs_structure}, the spiral arm pitch angles were compared for a number of different galaxy structural parameters. The first measure that was compared was the spiral arm number of the host galaxy, measured using the GZ2 visual morphologies. We saw that many-armed galaxies generally have looser spiral arms than their two-armed counterparts, irrespective of host galaxy stellar mass. Studies of local grand design spiral galaxies such as M51a show us that many two-arm structures are genuine density waves \citep{Colombo_14,Schinnerer_17}. Many-arm structures are instead usually considered to be modal structures arising in gas rich discs \citep{Carlberg_85,Baba_13}. Arm number vs. pitch angle correlations have been noted before, where a dependence on pitch angle on Elmegreen spiral arm class has been observed. \citet{Garcia_Gomez_93} noted that there is a correlation, where many-arm and flocculent spiral arms are looser than two-arm structures. Our results confirm the correlation of \citet{Garcia_Gomez_93} for a large statistically complete sample of spiral galaxies, with two-arm grand design spirals having tighter spiral arms than each of the many-arm categories, which we expect to include both many-arm and flocculent Elmegreen-type spirals.

A potential explanation for the differences in spiral arm pitch angles in different galaxy structures is due to the varying timescales over which spiral arms exist. Simulations of many-arm structures show that the pitch angle of individual spiral modes wind up over time (e.g. \citealt{Wada_11,Grand_12a,Grand_12b,Baba_13}). Many simulations also predict that these structures are usually short-lived phenomena and are usually broken or merged into other structures after $\sim$100Myr \citep{Baba_13,Donghia_13}, where spiral arms with $\psi>20$\textdegree{}  likely to be more transient features \citep{Perez_Villegas_12} than those with $\psi \leq$20\textdegree{}. However, the mechanisms responsible for two-arm structures are different from those in many-arm galaxies. The timescale over which two-arm structures can exist is still debatable, with some studies suggesting they are also transient phenomena \citep{Merrifield_06} but can potentially persist for longer when considering the gas component of galaxy discs \citep{Ghosh_15,Ghosh_16}. Two arm structures can also be tidally induced and wind up and decay over $\sim$1Gyr \citep{Oh_08,Struck_11}. It may therefore be the case that the tightly wound unbarred spirals we observe are the remnants of long-lived internal structures or the later stages of tidal features. 

Another reason for the observed differences in pitch angle with spiral arm number may be related to the rotation curves of galaxies. \citet{Seigar_05} demonstrated that spiral arm pitch angles can be directly related to the shear in the discs of galaxies: discs with falling rotation curves have tighter spiral arms than discs with rising rotation curves. Some tentative evidence that many-arm galaxies have steeper, rising rotation curves has been found before \citep{Biviano_91}, so these results may also indicate that many-arm structures arise in discs with lower shear rates. However, galaxy shear rates are closely related to the central mass concentrations in galaxies \citep{Seigar_05}, which are in turn related to spiral arm pitch angles \citep{Seigar_08}. We see no trend for galaxies with greater central mass concentrations having tighter spiral arm pitch angles, or more spiral arms as shown in Sec.~\ref{sec:results_mass}, suggesting that these differences are not driven by differences in the central mass concentrations.

\subsubsection{The role of bars}
\label{sec:discussion_bars}

In Sec.~\ref{sec:results_bars}, the \sparcfire{} measured spiral arm pitch angles are related to the presence of bars in the discs of galaxies, finding that galaxies with bars have looser spiral arms. This trend is particularly apparent when considering the two-arm population of spirals. Since grand design, two-arm structures are usually linked with both bars and companions \citep{Kormendy_79,Seigar_98a,Kendall_11}, and bars can be tidally induced \citep{Semczuk_17}, one possibility is that loose spiral arms in barred galaxies can potentially be the result of a galaxy-galaxy interaction disturbing the structure and forming loose tidal tails and bridges. Although bars are more common in higher density environments \citep{Skibba_12,Smethurst_17}, GZ2 statistics from \citet{Casteels_13} suggest that loose arms and bars are not due to pair interactions. As \citet{Casteels_13} showed that bars are actually suppressed, but the frequency of two-arm structures and loose spiral arms actually increases in close galaxy pairs, we do not favour this scenario. Looking for further evidence would require looking for interaction signatures at the most local galaxy scales, which are beyond the scope of this paper. 

A second possibility is that the presence of bars may have a strong influence on the dynamics within the discs of galaxies. Of particular interest is the `invariant manifold theory' \citep{Romero_Gomez_06, Romero_Gomez_07, Athanassoula_09a,Athanassoula_09b,Athanassoula_10}. A key prediction of this theory is that galaxies with stronger bars will have looser spiral arms than those with weak bars. Interpreting the GZ2 $p_\mathrm{bar}$ statistic as a relative measure of bar strength, our result would be in direct agreement with this prediction. We do, however, note that this evidence is somewhat tentative, as there are also other possibilities related to the dynamics of spiral galaxies that could give rise to this result. For example, \citet{Baba_15} describe a scenario where grand design spiral arms are amplified as they come closer to a bar, meaning that we may be seeing a signature that stronger bars amplify more efficiently at larger pitch angles. Another possibility is that the dynamics of spiral arms are altered when the bar becomes prominent \citep{Roca_Fabrega_13}. In this case, it has been suggested that the arms are better fit with a rigidly rotating disc, which would lead to arms which have hyperbolic rather than logarithmic patterns \citep{Seiden_79,Kennicutt_81}, potentially closing to ring-like features which cannot be described by a spiral equation with a single pitch angle \citep{Buta_86,Buta_17}. In this set of \sparcfire{} models, only logarithmic spirals are used to measure arms, so the differences in pitch angle could also be an artefact of mis-fitting the arms with a function that does not represent the spiral arm. If the logarithmic spiral equation was not the correct equation to fit, one would expect fewer galaxies to have detected arms. However, we see no difference in the number of galaxies with detected spiral arms with respect to $p_\mathrm{bar}$, meaning that spiral arms in barred galaxies are log spiral arcs, disfavouring the possibility that they are rigidly rotating arms.

\subsubsection{Galaxy mass concentrations}
\label{sec:mass_concentration}

In Sec.~\ref{sec:results_mass}, the mass properties of our unbarred galaxies were compared with respect to their spiral arm morphologies. Galaxies with more massive discs have more spiral arms, which in turn means that galaxies with lower bulge-to-total ratios have more spiral arms. However, we see no trend with respect to bulge mass. Given that pitch angle correlates with arm number, one naively would expect a positive correlation between arm number and disc mass, but we find a weak negative trend that galaxies with more massive discs and lower bulge-to-total ratios have looser spiral arms. The first result is perhaps surprising when considering simulations of modal spiral arms, which suggest that when galaxies have more massive bulges, higher order modes (more spiral arms) will dominate \citep{Grand_13,Donghia_15}. These models do usually consider isolated galaxies and some assumed dark matter halo profile. The dominant spiral arm mode will actually depend on the scale and mass of the dark matter halo as well as the mass of the disc and bulge, requiring a more complex analysis. Our results suggest that the spiral arms in our galaxies are not modal in nature, but driven by other processes, such as density waves or galaxy interactions. However, a more complete analysis of these models, including all of the relevant parameters is required to confirm this result, which will be done in a subsequent paper. This result would also appear to contradict the idea that galaxies with greater levels of shear have more tightly wound spiral arms \citep{Seigar_05,Seigar_08}. We note that these studies are usually done on small samples of nearby galaxies, with clear, unbarred grand design spiral arms. Our result that bulge mass has no influence on arm pitch angle suggests that this relation does not hold for the entire spiral galaxy population, and that spiral arm pitch angle is more heavily influenced by other properties, such as disc mass, arm number and the potential presence of a bar, rather than just central mass concentration.

\subsection{Star formation rates}
\label{sec:discuss_SFRs}

In Sec.~\ref{sec:SFRs}, it was shown that SFR depends on spiral arm number. Galaxies with more spiral arms tend to have more UV and less mid-IR emission, indicating that the SFR is less obscured by dust in many-armed galaxies, but that the overall SFRs are similar in all types of spiral galaxies. The reasons for these trends are discussed in more detail in \citet{Hart_17}, but can be related to the geometry of star formation and molecular clouds in galaxies. We also found that SFRs are consistent in spiral galaxies irrespective of arm number, other than when the number of spiral arms was less than two. We saw an increase in both the mid-IR and total SFR. Given that one-arm galaxies in GZ2 have been associated with merger remnants \citep{Casteels_13}, this would suggest that high SFRs are triggered by galaxy-galaxy interactions \citep{Barton_00,Ellison_08,Willett_15}. Despite the strong, clear trends with spiral arm number, we found no differences in the SFRs with spiral arm pitch angle. 
In grand design spiral galaxies, shear rates are related to spiral arm pitch angles \citep{Seigar_08,Seigar_14}. If the shear is too high, galaxies have been shown to have lower total SFRs \citep{Seigar_05}. One would therefore expect galaxies with tighter spiral arms to have lower total SFRs if shear was responsible for the differences in spiral galaxy pitch angle. Given that we find no relation between SFR and pitch angle, this would suggest that shear is not reposnsible for pitch angle differences between galaxies. We also find no enhancements in the SFRs of spirals with looser arms, meaning that the loosest arm spirals are unlikely to be merger remnants like one-arm spirals. 

\section{Conclusions}
\label{sec:conclusions}

In this paper, the spiral arm pitch angles for a sample of GZ2 identified spiral galaxies were measured and compared. The arm pitch angles in multi-arm spiral galaxies are on average larger than those in grand design, two-arm spirals. We suggest a number of possible reasons for these results, including the ideas that multi-arm patterns may be short-lived or that the mechanisms that cause galaxies to have different arm numbers also influence the spiral arm pitch angles. We also find evidence that spiral arms are looser when strong bars are present in galaxies. This can be interpreted as evidence for bars influencing the dynamics of galaxies, such as in invariant manifold theory, or as evidence that barred galaxies are potentially triggered by interactions. We showed that many-arm structures are more prevalent in galaxies with less dominant bulges and that pitch angles are looser when the bulge is more dominant. These results imply that the spiral galaxies that we observe are not truly modal in nature and that the shear rates in discs are not the dominant reason for variations in spiral arm pitch angle. Finally, we saw weak trends that galaxies with more mid-IR star formation have looser spiral arms, an effect that could be related to shear or galaxy-galaxy interactions.

Given that spiral arms are fundamental features seen in significant numbers of low-redshift galaxies, the results in this paper show that their origins are still not well-understood. It has been fifty years since the density wave theory was proposed \citep{Lindblad_63,Lin_64} as the principal reason for the appearance of spiral arms in galaxies and thirty years since spiral arms have been formed in simulations \citep{Sellwood_84}. With this detailed investigation into the number of arms formed and their tightness we are able to show how complex spiral formation patterns are and that some of the most basic scaling relations do not seem to hold for the majority of low-redshift spirals. Given that bar and disc properties seem to be as important as central mass concentrations in influencing the pitch angles of spiral galaxies, there are a multitude of complex mechanisms at play in galaxies that affect the properties of spiral arms. The use of methods like those described in this paper to detect individual spiral arms in galaxies, coupled with a wealth of dynamical data becoming available \citep{Bryant_15,Bundy_15}, can potentially help to disentangle the different mechanisms that affect the spiral arm morphology in galaxies.

\section*{Acknowledgements}

The data in this paper are the result of the efforts of the Galaxy Zoo 2 volunteers, without whom none of this work would be possible. Their efforts are individually acknowledged at \url{http://authors.galaxyzoo.org}.

The development of Galaxy Zoo was supported in part by the Alfred P. Sloan foundation and the Leverhulme Trust.

REH acknowledges a studentship from the Science and Technology Funding Council, and support from a Royal Astronomical Society grant. RJS gratefully acknowledges funding from the Ogden trust.

This paper makes extensive use of \texttt{scikit-learn} \citep{scikit-learn} and \texttt{astroML} \citep{astroML}. This publication also made extensive use of the \texttt{scipy} Python module \citep{scipy}, \texttt{TOPCAT} \citep{topcat} and \texttt{Astropy}, a community-developed core Python package for Astronomy \citep{astropy}. We acknowledge the work of the people involved in the \textsc{Zooniverse} project-builder for the construction of the \spotter{} project.

Funding for SDSS-III has been provided by the Alfred P. Sloan Foundation, the Participating Institutions, the National Science Foundation, and the U.S. Department of Energy Office of Science. The SDSS-III web site is http://www.sdss3.org/.

SDSS-III is managed by the Astrophysical Research Consortium for the Participating Institutions of the SDSS-III Collaboration including the University of Arizona, the Brazilian Participation Group, Brookhaven National Laboratory, Carnegie Mellon University, University of Florida, the French Participation Group, the German Participation Group, Harvard University, the Instituto de Astrofisica de Canarias, the Michigan State/Notre Dame/JINA Participation Group, Johns Hopkins University, Lawrence Berkeley National Laboratory, Max Planck Institute for Astrophysics, Max Planck Institute for Extraterrestrial Physics, New Mexico State University, New York University, Ohio State University, Pennsylvania State University, University of Portsmouth, Princeton University, the Spanish Participation Group, University of Tokyo, University of Utah, Vanderbilt University, University of Virginia, University of Washington, and Yale University.




\bibliographystyle{mnras}
\bibliography{bibliography} 



\appendix
\section{Support Vector Machine arm classification}
\label{appendix:SVM}

Given the statistics from \spotter{}, the overall characteristics of true arms are assessed. The resulting distributions are shown in Fig.~\ref{fig:svm_histograms}. From these distributions, true spiral arms can be identified by the following characteristics:

\begin{figure}
    \includegraphics[width=0.45\textwidth]{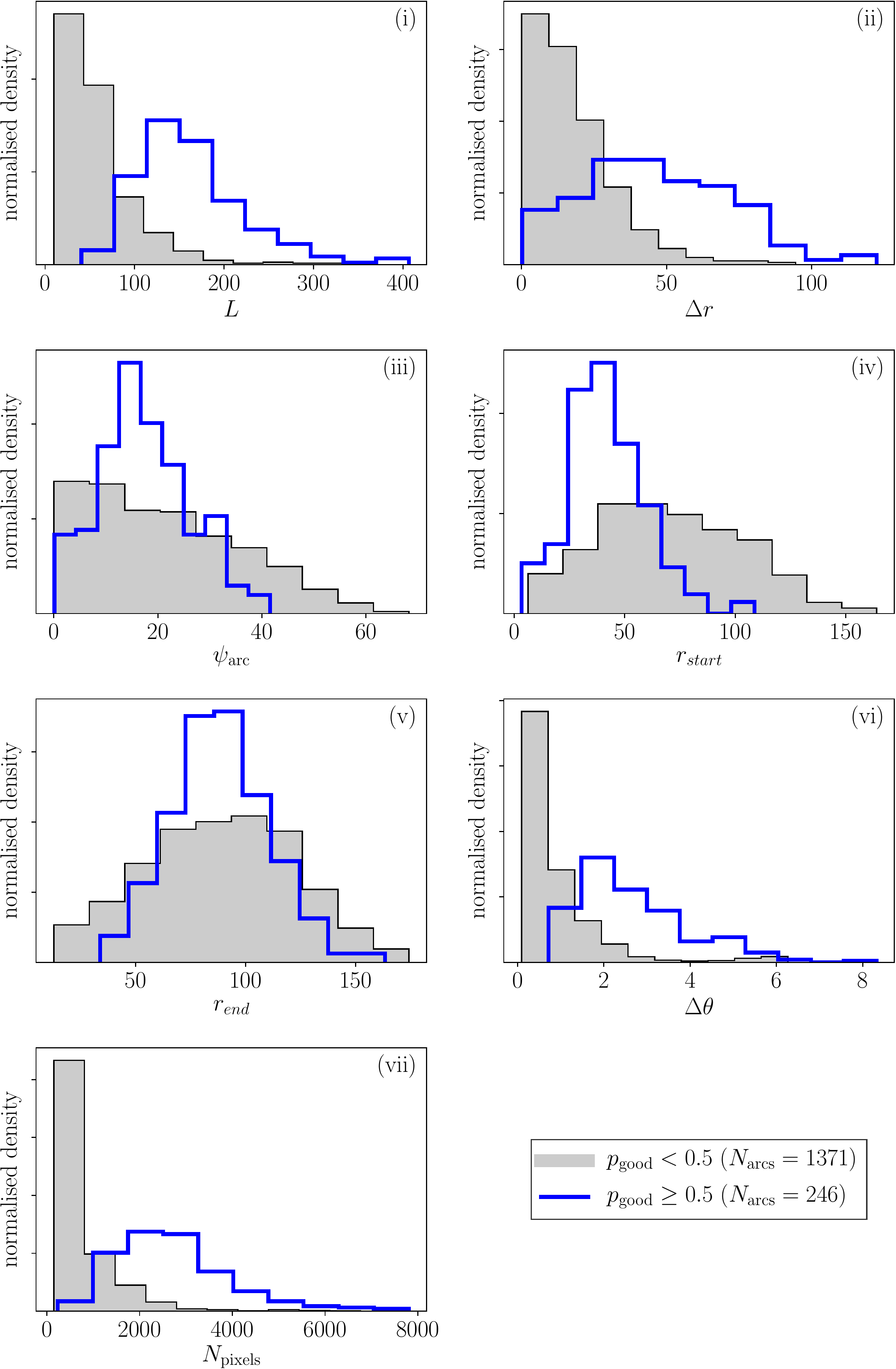}
    \caption{Distributions of the seven parameters listed in appendix ~\ref{appendix:SVM} for arcs that have been visually identified as poor arcs ($p_\mathrm{good}<0.5$, grey filled histograms) and good fits to real spiral arms ($p_\mathrm{good} \geq 0.5$, blue stepped histograms).}
    \label{fig:svm_histograms}
\end{figure}

\begin{enumerate}

\item $L$ (arc length): this is the primary way in which true spiral arcs have been distinguished from noise in \citet{Davis_14}. Generally, we expect that longer arcs are more likely to be real objects. Our \spotter{} analysis shows that this is indeed the case.

\item $\Delta r$ (radial arm range): generally, we see that true spiral arcs are more likely to cover a wider range of the galaxy's radius.

\item $\psi_\mathrm{arc}$ (arc pitch angle): true spiral arcs seem to preferentially occupy the range $10 < \psi_\mathrm{arc} < 40$\textdegree. This is similar to the range observed in other samples of nearby galaxies (e.g. \citealt{Seigar_08}).

\item $r_\mathrm{start}$ (initial arc radius): generally, true spiral arcs are more likely emanate from closer to the centre of galaxies.

\item $r_\mathrm{end}$ (end arc radius): this parameter appears to have little influence, but we do see that true spiral arcs tend to end at $\sim$100 pixels. We note that \sparcfire{} scales all images using an isophotal ellipse fitting routine, so these distances are similar in all galaxies.

\item $\Delta \theta$ (angular extent of the arms): true spiral arms have longer $\Delta \theta$ values, meaning that they wrap further around the centre of the host galaxy.

\item $N_\mathrm{pixels}$ (number of pixels that the \sparcfire{} arc mask comprises): true spiral are drawn through regions made up of more pixels.

\end{enumerate}

We can therefore look for these characteristics in the spiral arms of all of our galaxies. However, these data form a high-dimensional space, with multiple underlying correlations between the variables. Rather than define individual data cuts, we instead train a machine learning classifier on these arc characteristics. After evaluating all possibilities, we elect to use a called a support vector machine or SVM, using the \textsc{svm.svc} package from \textsc{scikit-learn} \citep{sklearn}. The SVM was trained on all seven of the variables listed above.

\begin{figure}
    \includegraphics[width=0.45\textwidth]{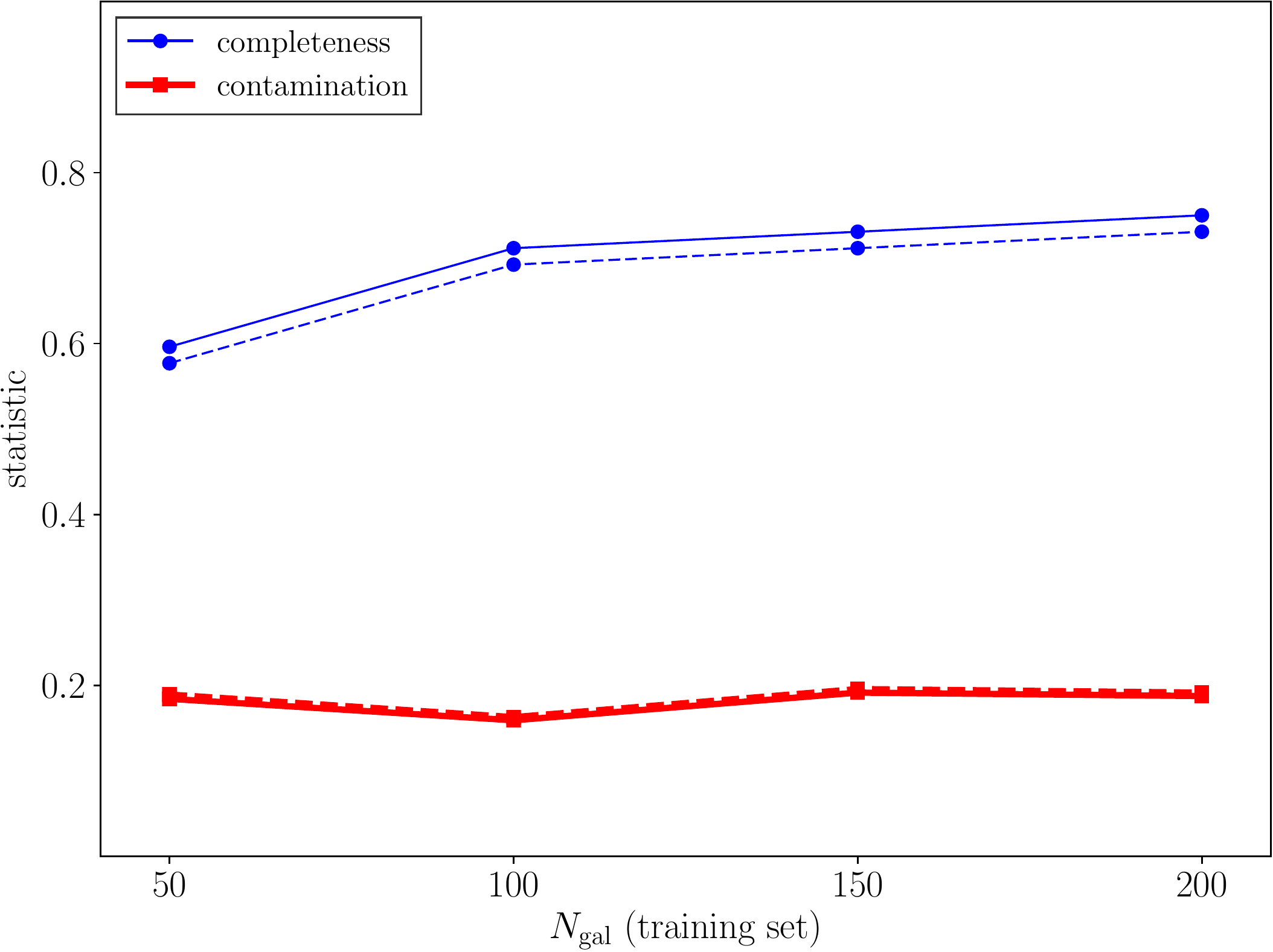}
    \caption{Completeness (thin blue line) and contamination (thick red line) of the test set of galaxies, for the SVM trained on samples of different sizes. The solid lines indicate the the statistics when all spiral arcs are considered, and the dashed line shows the same statistics when only arcs with agreement with the dominant galaxy chirality (discussed in Sec.~\ref{sec:identifying_arms}) are considered.}
    \label{fig:svm_performance}
\end{figure}

From \spotter{}, we have 252 galaxies which have been visually inspected. Training our SVM requires a training set and an independent test set to check its results. The training set was a randomly selected set of 200 galaxies and the test set was made up of the remaining 52 galaxies (an approximately 80:20 split). The test set was kept separate from the training set and the SVM was trained on four subsamples, including 50, 100, 150 and 200 galaxies respectively, to check its performance as more data was included. The completeness and contamination for the SVM trained on each of the subsamples are shown in Fig.~\ref{fig:svm_performance}. We see that including more galaxies delivers a marginal improvement in terms of completeness as more galaxies are trained upon, and the contamination stays relatively constant. Our SVM is therefore trained on these 200 galaxies, as adding in more galaxies does little to improve the performance of the SVM classifier. This trained SVM is then applied to the full sample of galaxies discussed in Sec.~\ref{sec:sample_selection} and later, to identify reliable logarithmic arcs in galaxies.


\bsp	
\label{lastpage}
\end{document}